\documentclass[preprint,1p]{elsarticle}
\usepackage{etex}
\usepackage[T1]{fontenc}
\usepackage[utf8]{inputenc}
\usepackage[english]{babel}
\usepackage{multirow}
\usepackage{graphicx}
\usepackage[dvipsnames]{xcolor}
\usepackage{float}
\usepackage{enumitem}
\usepackage{caption}
\usepackage{amssymb,amsfonts,amsmath,mathtools}
\usepackage[all,pdf,cmtip]{xy}
\usepackage{multicol}
\usepackage[scaled=0.9]{helvet}
\usepackage[colorinlistoftodos]{todonotes}
\usepackage{hyperref}

\newcommand{\Typed}[1]{\ensuremath{#1\textrm{-}\mathbf{Graph}}}
\newcommand{\mc}[1]{\ensuremath{\mathcal{#1}}}

\usepackage{tikz}

\renewcommand{\urcorner}{{\tikz \draw[-] (5pt,0)   -- (5pt,5pt) -- (0,5pt);}}
\renewcommand{\ulcorner}{{\tikz \draw[-] (5pt,5pt) -- (0,5pt)   -- (0,0);}}

\newenvironment{example}{ Example: \it}{ \rm}

\usepackage{amsthm}
\newtheorem{defn}{Definition}
\newtheorem{theorem}{Theorem}

\newtheorem{exmp}{Example}
\newcommand{\nac}{N\!AC}
\newcommand{\Span}{\ensuremath{\mathbf{Span}}}

\usepackage{tikz,tkz-euclide}
\usetikzlibrary{backgrounds}
\usepackage[all,pdf,cmtip]{xy}
\usepackage{tikz-cd}
\usepackage{framed}

\journal{Science of Computer Programming}

\begin{document}

\begin{frontmatter}

\title{Use Case Evolution Analysis based on Graph Transformation with Negative Application Conditions \thanks{This work is partially supported by the VeriTeS project (CNPq).}}
\author{Leila Ribeiro \corref{mycorrespondingauthor}}
\author{Lucio Duarte}
\author{Rodrigo Machado}
\author{Andrei Costa}
\author{Érika Cota}
\author{Jonas Bezerra}
\address{{PPGC/INF - Federal University of Rio Grande do Sul (UFRGS) -- 
Porto Alegre, Brazil\\
Email: \{leila, lmduarte, rma, acosta, erika, jsbezerra\}@inf.ufrgs.br}  
}

\begin{abstract}
\emph{Use Case (UC)} quality impacts the overall quality and defect rate of a system, as they specify the expected behavior of an implementation. In a previous work, we have defined an approach for a step-by-step translation from UCs written in natural language to a formal description in terms of Graph Transformation (GT), where each step of the UC was translated to a transformation rule. This UC formalisation enables the detection of several specification problems even before an actual implementation is produced, thus reducing development costs. In this paper, we extend our approach to handle UC evolution by defining \emph{evolution rules}, which are described as higher-order rules, simultaneously changing the behaviour of a set of transformation rules. We also support the use of \emph{negative application conditions (NAC)} associated both to the transformation and evolution rules. Analysis of the interplay between the evolution rules and the rules describing UC steps shows the effects of an evolution and serves to identify potential impacts, even before the changes are actually carried out. Besides defining the theoretical foundations of UC evolution with NACs, we have implemented the evolution analysis technique in the Verigraph tool and used it to verify impacts in 3 different case studies. The results demonstrate the applicability and usefulness of our approach to help developers in the evolution process based on UCs.
\end{abstract}

\begin{keyword}
Use Case \sep Graph Transformation \sep Software evolution \sep Higher-order transformations \sep NACs
\end{keyword}

\end{frontmatter}

\section{Introduction}
\label{sec:introduction}

Use Cases (UC) descriptions are typically informally documented, using natural language \cite{Cockburn2000}. Being informal descriptions, UCs might be ambiguous and imprecise, resulting in a number of problems that can propagate to later development phases and jeopardize the overall system quality \cite{Alagar2011}. Effective manual verification of these artifacts may be expensive, which makes it desirable to have a way of automating this task. UC correctness is specially vulnerable during system evolution. As a system evolves, specification documents must change accordingly, describing new functionalities and/or including changes in existing features. Depending on the number of UCs and their complexity, identifying the impacts of an evolution can be extremely hard to do by human inspection.
Ensuring that the UCs have been correctly updated to reflect the necessary modifications can be a costly and time-consuming task. Hence, this process should be supported by techniques that could guide the developer, indicating how to apply each of the required changes to the UCs and how these changes  affect local (intra-UC) and global (inter-UC) behaviour. Thus, the translation of UCs to a formal model can help improve UC reliability and correctness by enabling a variety of analyses using existing tool support. This allows the detection of several specification problems before an actual implementation is produced, thus reducing development costs. 

 


In \cite{LNWADT2015}, we presented an approach for the translation of UCs to a formal model called \emph{Graph Transformation (GT)}. Elements of a system are described as nodes of a graph and edges between them represent their relationships (e.g., data dependency). System behavior is specified by a set of transformation rules, creating/preserving/deleting nodes and/or edges. Each rule specifies application pre-conditions (i.e., the necessary set of nodes and their specific connections through edges that must exist as a subgraph of the current system graph) and post-conditions (i.e., the resulting subgraph). Besides being a \emph{visual} and \emph{intuitive notation}, GT is \emph{data-driven}, allowing a very abstract description of a system, without imposing any control not strictly enforced by data. Moreover, GT analysis is supported by tools capable of pinpointing possible problems, which are easily traced back to the original UC description. 

The proposed translation method, although not yet fully supported by tools, can be followed by developers with very basic knowledge of the formalism, as we provide step-by-step guidelines. The subsequent analyses are automatically performed by available tools (AGG \cite{Taentzer2000} and Verigraph\footnote{\url{http://github.com/verites/verigraph}}). Types and severity of different classes of problems that can be detected have been defined, as well as hints on the causes and possible solutions. The UCs can then be modified until the analyses show that they present the desired behavior. As a result, the GT-based UC verification approach increases the correctness of the UCs, while keeping the informality and flexibility of their description in natural language. UC analysis based on GTs has been evaluated on real UCs and revealed several problems not detected during manual inspection \cite{LNWADT2015}. From these problems, 75\% were classified as actual errors by the developers who had created the UCs. Hence, the approach has proved to be useful and effective for UC analysis and enhancement.

In this paper, \emph{we extend our approach to support the description and analysis of system evolution}. By "evolution" we mean any modification in the system, which may or may not affect the system behaviour. We expand the formal framework to define evolution in terms of higher-order GTs, introduced in \cite{Machado2015}. The main idea is to describe evolution as a second-order rule, called \emph{evolution rules}, which are rules that change other (first-order) rules. Because evolution is also represented as a GT, the developer works within the same formalism used to specify the system. 
We present the theoretical foundations for reliable UC change in a system evolution scenario, offering not only a precise description of evolution, but also the possibility of analyzing evolution effects even before modifying the original rules, thus helping the decision-making process. The analysis technique is based on critical pairs, and was implemented in the \emph{Verigraph} tool, which is also able to apply the changes defined by an evolution rule. We also extend the framework of higher-order graph transformation to support evolution of rules with \emph{negative application conditions (NACs)}, since NACs  are required in the GT-based representation of use cases.
Moreover, we provide guidelines to interpret the analysis results and demonstrate the application of our evolution framework in 3 case studies.

This paper contains the following structure: Section~\ref{sec:background} presents the background on GTs, the UCs formalization strategy proposed in \cite{LNWADT2015}, and critical pair analysis in the context of UCs; 
Section~\ref{sec:sogt-nacs} extends the notion of second-order transformation towards rules with negative application conditions (NACs).
Section~\ref{sec:hogt} presents our approach to describe and analyze UC evolution; in Sect.~\ref{sec:experiments} the application of the proposed approach on 3 case studies is presented; Section~\ref{sec:relatedwork} describes some related work; and Section~\ref{sec:conclusions} presents conclusions and possible future work.

\section{Background} 
\label{sec:background}

\subsection{Graph Transformations}
\label{sec:gt}

In this section, we present the main concepts of Graph Transformations (GT). The algebraic approaches for graph transformation  use categorical operations in order to perform the transformations  defined by the rules~\cite{ehrig1973,ehrig1978}. The approach  we follow  uses two \emph{pushouts} as gluing operations, therefore it is called Double-PushOut approach (DPO). Examples are presented in the next section.


\textbf{Graphs} are structures that consist of a set of nodes and a set of edges. Each edge connects two nodes of the graph, one representing a source and another representing a target. A \textbf{(total) homomorphism} between graphs is a mapping of nodes and edges that is compatible with sources and targets of edges. Intuitively, a  homomorphism from a graph \textit{G1} to a graph \textit{G2} means that all items (nodes and edges) of \textit{G1} can be found in \textit{G2} (but distinct nodes/edges of \textit{G1} are not necessarily distinct in \textit{G2}). 

\begin{defn}[graph, graph morphism]
A \textbf{graph} $G = (V,E,s,t)$ consists of a set $V$ of nodes, a set $E$ of edges, and two functions, $s,t:E \rightarrow V$, the source and target functions. Given two graphs, $G_1 = (V_1,E_1,s_1,t_1)$ and $G_2 = (V_2,E_2,s_2,t_2)$, a \textbf{graph morphism} $f:G_1 \rightarrow G_2$ is composed by two total functions $f_V:V_1 \rightarrow V_2$ and $f_E:E_1 \rightarrow E_2$ such that $f_V \circ s_1 = s_2 \circ f_E$ and $f_V \circ t_1 = t_2 \circ f_E$. A  graph morphism is injective/surjective/iso if both components are injective/surjective/iso. The category of graphs and graph morphisms is called $\mathbf{Graph}$.
\end{defn}

For practical applications, it is very convenient to distinguish different types of vertices and edges in a graph. In the DPO approach, this can be achieved by the notion of typed graph. Let \textit{TG} be a graph that represents all possible (graphical) types that are needed to describe a system, a homomorphism \textit{h} from any graph \textit{G} to \textit{TG} associates a (graphical) type to each item of \textit{G}. The triple $\langle$\textit{G, h, TG}$\rangle$ is called  \textit{typed graph}, where \textit{TG} is the \textit{type graph} (nodes of \textit{TG} denote all possible types of nodes of a system, edges of \textit{TG} denote possible relationships between the nodes).

\begin{defn}[typed graph, typed graph  morphism]
A \textbf{typed graph} is a triple $(G1,type_{G1}, TG)$, denoted by $G1^{TG}$, where $G1$ and $TG$ are graphs and $type: G1\rightarrow TG$ is a graph morphism.
Given two typed graphs over the same type graph, $G_1^{TG}$ and $G_2^{TG}$, and their respective typing morphisms $type_{G1} : G_1 \rightarrow TG$ and $type_{G2} : G_2 \rightarrow TG$, a 
\textbf{typed graph morphism} is a pair $(f,id_{TG})$, where $f : G_1 \rightarrow G_2$ is a graph morphism and $id_{TG}$ is the identity morphism of $TG$, such that: $type_{G2} \circ f = type_{G1}$.
A typed graph morphism is injective/surjective/iso if $f$ is injective/surjective/iso.

\centerline{\xymatrix{
G_1\ar[dr]_{type_1}\ar[rr]^{f} &\ar@{}[d]|{=} & G_2\ar[dl]^{type_2} \\
 & TG & }}

The category of graphs typed over $TG$ as objects and typed graph morphisms as morphisms is called $\mathbf{Graph}_{TG}$. This category is the comma category $(\mathbf{Graph}\downarrow \mathbf{TG})$.
\end{defn}

A \textbf{Graph Rule} describes how a system may change. It consists of: a \textbf{left-hand side (LHS)}, denoting items that must be present for this rule to be applied; 
a \textbf{preserved part (also called gluing part K)},
describing  items that will be preserved when the rule is applied; a \textbf{right-hand side (RHS)}, describing items that will be present after the application of the rule;  \textbf{mappings from K to LHS and RHS}, which mark in LHS and RHS the items that will be preserved by the application of the rule; and a \textbf{negative application condition (NAC)}, that is actually a collection of conditions representing situations that prevent the rule from being applied (NACs are described by mappings from LHS to the graph representing the forbidden context). The mappings from K to LHS and RHS (as well as from LHS to the NACs) must be compatible with the structure of the graphs (graph homomorphisms). Items that are in LHS and are not in K are \textbf{deleted}, whereas items that are in the RHS and are not in  K are \textbf{created}. We assume that rules do not merge items (rules are injective). The  structure $LHS \leftarrow K \to RHS$  of a rule is called \textbf{rule span}.

\begin{defn}[rule, NAC, rule with NACs, NAC satisfiability]\label{def:nac}
A (typed graph) rule $p$ consists of two typed graph morphisms $l$ and $r$ with the same typed graph as source, $p = L \xleftarrow{l} K \xrightarrow{r} R$, where $l$ and $r$ are monomorphisms (in the category  $\mathbf{Graph}_{TG}$, monomorphisms are injective mappings).

Given a rule  $p = L \xleftarrow{l} K \xrightarrow{r} R$, a \textbf{negative application condition} $nac(n)$ for $p$  is an arbitrary typed graph morphism $n : L \rightarrow N$. A NAC $n : L \rightarrow N$ is satisfied with respect to a match $m : L \rightarrow G$  if and only if $\nexists q : N \rightarrow G$ such that $q$ is injective and $q \circ n = g$.

A rule with NACs $(p,nac_p)$ is composed by a rule $p$ and a set of NACs for $p$ ($nac_p$).

A match $m : L \rightarrow G$ satisfies $nac_p$ if and only if it satisfies all single NACs in $nac_p$, we denote as $nac_p \vDash m \left(p,G\right)$.

\centerline{\xymatrix{
~ & N_i\ar@{>.>}[dl]_q \\
G & L\ar[l]^m\ar[u]_{n_i}}}
\end{defn}

A \textbf{GT System} consists of a type graph, specifying the (graphical) types of the system, and a set of rules over this type graph that define the system behavior. 

\begin{defn}[(first order) graph transformation system]
A \textbf{(first order) Graph Transformation System}   consists of a set of (typed graph) rules with NACs and a  typed graph, called initial or start graph. 
The rules of a  first-order graph transformation system are called  first-order rules.
\end{defn}

The \textit{application of a rule} $(NAC,rs)$ where $rs: L \leftarrow K \to R$  to a graph $G$ is possible if (i) an image of $L$  is found in $G$ (that is, there is a total typed-graph morphism from the LHS of $rs$ to $G$); (ii) for each $nac: L \to X$ in $NAC$ it is not possible to find an image for $X$ in $G$ (i.e., the forbidden items are not in $G$); and
(iii) the \textbf{gluing condition} is satisfied, this  condition assures that the result of removing from $G$ all items that should be deleted by the rule yields a unique result that is a well-formed graph (for example, it is not possible to delete a node from $G$ if there are edges connected to it that are not also marked for deletion in $L$).
The result of a rule application deletes from $G$ all items that are not in the gluing graph $K$ (step (1) below) and adds the ones created in $R$ (step 2 below). Formally, these steps are constructed by pushouts in a suitable category.

\[
\xymatrix@=1cm{ 
*++{X} \ar@{>-->}@/_0.7cm/[dr]|{\diagup} \ar@{>-->}@/_0.7cm/[dr]|{\diagdown} \ar@{>-->}@/_0.7cm/[dr]_(.3){e}   &     *++{L} \ar[l]_{n_i}  \ar@{->}[d]_{m}
    & *++{K}  \ar@{}[dl]|{(1)} \ar@{}[dr]|{(2)}  \ar@{>->}[l]_{l} \ar@{>->}[r]^{r} \ar[d]^{} 
    & *++{R} \ar[d]^{}
\\  
&     *++{G}
    & *++{D} \ar[l]^{} \ar[r]_{}
    & *++{H} \\
}
\]

\begin{defn}[match, typed graph transformation]
Consider a rule $p = L \xleftarrow{l} K \xrightarrow{r} R$ and a typed graph $G$, as in the diagram below.
A \textbf{match} is an arbitrary typed graph morphism from $L$ to $G$.

A match $m$ satisfies the \textbf{gluing condition} iff both conditions below are satisfied:
\begin{description}
\item [(dangling condition)] All edges of $G$ that are connected to nodes that are in the image of $m$ are also in the image of $m$;
\item [(identification condition)] If an element $e$ of $L$ is deleted (not in the image of $l$), no other element of $L$ may be mapped to $m(e)$ in $G$.
\end{description}

A match $m : L \rightarrow G$ satisfies $nac_p$ if and only if it satisfies all single NACs in $nac_p$, we denote as $nac_p \vDash m \left(p,G\right)$.

\centerline{\xymatrix{
~ & N_i\ar@{>.>}[dl]_q \\
G & L\ar[l]^m\ar[u]_{n_i}}}

Given a rule $p$ and a match $m$,  a \textbf{(typed) graph transformation} $G \xRightarrow{p,m} H$ from $G$ to $H$ is defined as the diagram below, where (1) and (2) are \emph{pushouts} in the category $\mathbf{Graph}_{TG}$. 

\centerline{\xymatrix{
L\ar@{}|{(1)}[dr]\ar[d]_m & K\ar@{>->}[l]_l\ar@{>->}[r]^r\ar[d] & R\ar@{}|{(2)}[dl]\ar[d]^{m'} \\
G\ar@{}[ur]|<{\urcorner} & D\ar@{>->}[l]^{l'}\ar@{>->}[r]_{r'} & H\ar@{}[ul]|<{\ulcorner}}}
\end{defn}



\subsection{UC Formalization} \label{sec:methodology}
\begin{figure}[!ht]
\centering
  \includegraphics[scale=0.3]{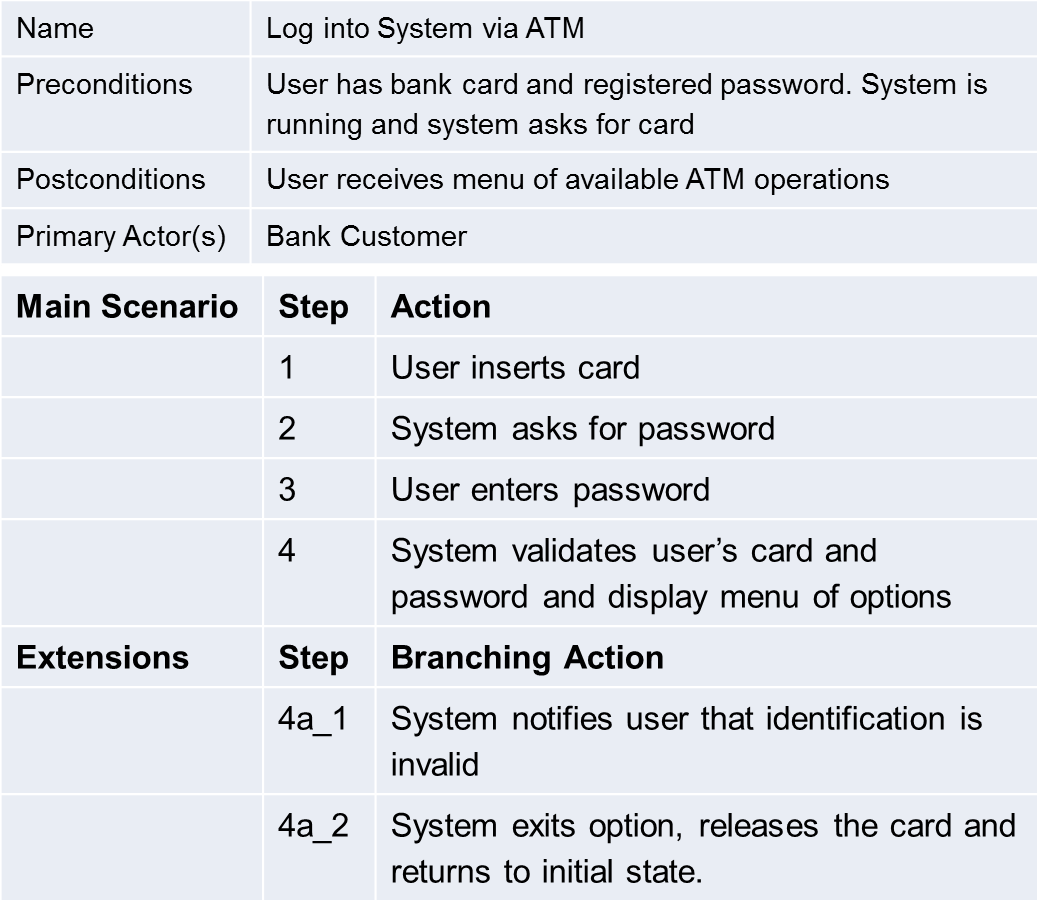}
  \caption{Login Use Case description.}
  \label{fig:UCLogin}
\end{figure}

The main purposes of a UC description are the documentation of the expected system behavior and the communication between stakeholders - often including non-technical people - about required system functionalities. For this reason, the most usual UC description is textual, using natural language. 
Figure~\ref{fig:UCLogin} depicts a UC of a bank system describing the login operation, executed by a client. 

We use a general UC description format, containing a primary actor and a set of sequential steps describing interactions between the primary actor and the system towards a certain goal. A sequence of alternative steps is often included to represent exception flows (\emph{Extensions} in Figure~\ref{fig:UCLogin}), when the primary goal is not achieved, or to describe alternative ways to achieve it. Pre- and post-conditions are also listed to indicate, respectively, conditions that must hold before and after the UC execution.


\begin{figure}[!ht]
\centering
\includegraphics[scale=0.4]{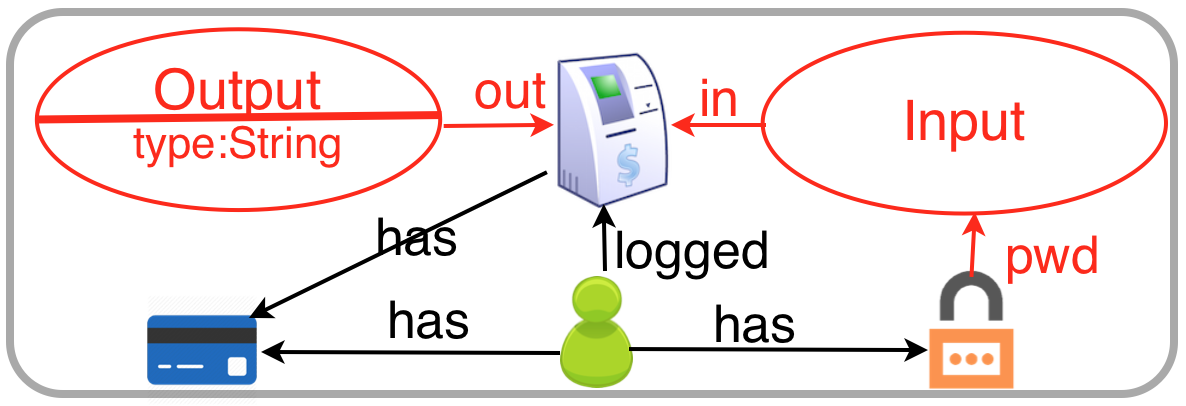}
\caption{Type graph.}
\label{fig:TypeGraph}
\end{figure}

In \cite{LNWADT2015}, we proposed a UC formalization and verification approach based on the GT formalism. 
Given a UC, we build a \emph{type graph} containing all relevant types for this UC and a set of rules describing its steps, one rule for each step, which makes it easy to understand the generated model and to trace it back to the UC. Figure~\ref{fig:TypeGraph} presents the type graph for our bank example: a card can be either in possession of a user or inside an ATM; a user can be logged to an ATM and has a password for access purposes, which is provided to the ATM via an input; the ATM can generate outputs to a user via a textual message.

\begin{figure}[htbp]
\vspace{-0.5cm}
\centering
\includegraphics[width=\textwidth]{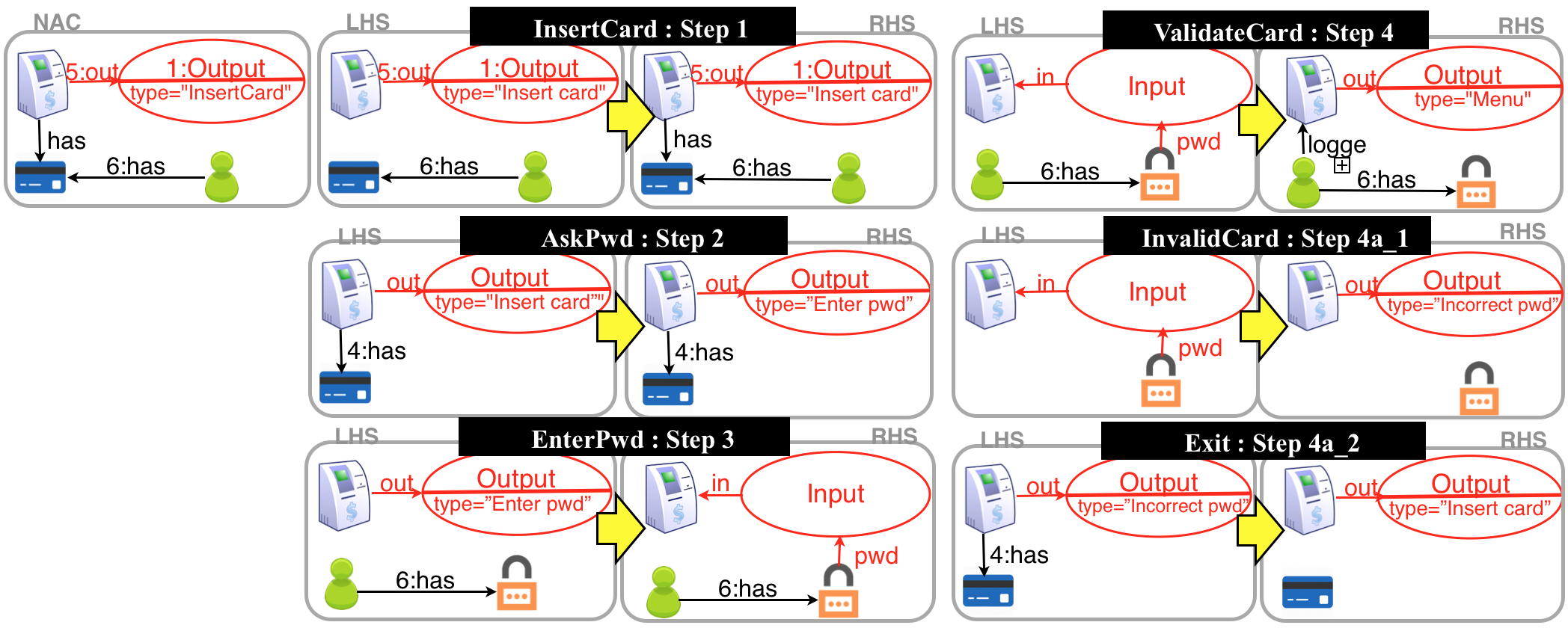}
\caption{Graph rules obtained from the individual steps of the example Use Case (Figure~\ref{fig:UCLogin}).}
\label{fig:gg-from-uc}
\vspace{-0.3cm}
\end{figure}

Figure~\ref{fig:gg-from-uc} presents the graph transformation rules derived from the \emph{Login Use Case}, shown in Figure~\ref{fig:UCLogin}. Each rule provides a representation of a step of the UC  (see labels inside black boxes). For instance, the act of a user inserting a card into an ATM is represented by the creation of an edge between the ATM and the card (rule \textsf{InsertCard}). The negative application condition (NAC) of rule \textsf{InsertCard} is used to disable the insertion of a card if there is already a card inside the ATM. The \textsf{Output} node  represents a message on the ATM screen, which is used to interact with a user. Its type indicates the type of the displayed message. 
The gluing component of the rule ($K$) is not explicitly represented; rather it is composed by items that are both in the LHS and the RHS of the rule (in the edges and Input/Output nodes, numbers are used to make the mapping more explicit).
Similarly, the \textsf{Input} node is used to represent data input. For example, rule \textsf{AskPwd} changes the output node type in the presence of a card. Then, rule \textsf{EnterPwd} represents the act of a user entering a password. From this point on, there are two possibilities: either the user enters the right password, and the system creates a user session (rule \textsf{ValidateCard}), or the password is invalid and the system presents a message (rule \textsf{InvalidCard}), and then ejects the card (rule \textsf{Exit}). 

After obtaining a first version of the GT from an UC, a series of automatic verifications can be performed using the AGG tool \cite{Taentzer2000}. All detected issues are annotated as \emph{open issues (OIs)} along with possible solutions (when applicable). Hence, any design decision made over an OI can be documented and tracked back to the original UC. One important point is that, during the formalization process, clarifications and decisions about the intended semantics of the textual description must be made. Annotated OIs force stakeholders to be more precise and explicit about tacit knowledge and unexpressed assumptions about system invariants and desired behavior. The result of applying the approach proposed in \cite{LNWADT2015} is an improved UC, as well as a GT representing the UC described behavior. This GT is the starting point of our contribution presented here: we propose a formal definition of evolution and an analysis method that shows the impact of this evolution on UCs (without actually performing the evolution). The analysis method is based on finding critical pairs between evolution rules and rules describing UC steps. We now review the main concepts of critical pair analysis in the context of GT.

\subsection{Use Case Analysis based on Critical Pairs} \label{sec:uca}

In GT, conflicts and dependencies between rules can be detected by computing \emph{critical pairs}, which are pairs of rules such that the application of one rule may have an impact on the application of the other (enabling or disabling it). For example, if a rule  $r_1$ deletes an item of type $X$ and a rule $r_2$ needs this item to be applied, then applying $r_1$ may prevent the application of $r_2$. Notice that this is a \emph{potential} conflict: if there are multiple items of type $X$ in the state graph, then the two rules may be applied independently, using different instances of $X$. There are essentially two types of conflict that may arise in GT systems with NACs  \cite{Lambers2006}:

\begin{description}
\item [Use-delete (ud):] one rule uses (deletes or preserves) an element that is deleted by another rule; 
\item [Produce-forbid (pf):] one rule produces an element that triggers some NAC of another rule. 
\end{description}

A pair of rule applications may have a conflict in just one direction ($r_1$ is in conflict with $r_2$, but $r_2$ is not in conflict with $r_1$), 
or in both directions, i.e., each rule deletes items that the other rule uses, or  creates items that trigger a NAC of the other rule. 
We refer to the latter cases, respectively, as \textbf{delete-delete (dd)} conflicts and \textbf{double-produce-forbid (ff)} conflicts.
Note that, since rules may delete/create/preserve many items, two rules may be simultaneously in more than one type of conflict.
Although there may be infinite conflicting situations involving two rules (because conflicts depend on the actual state in which rules are applied and there is usually an infinite set of states), 
there are only a finite number of critical pairs, which makes them useful for static analysis techniques.
Computing critical pairs involve comparing types of items created/preserved/deleted/forbidden by two rules. Since the comparison is based on types, \emph{critical pair analysis (CPA)} detects potential conflicts between rules.
For the formal definition of critical pairs see \cite{Lambers2006}.

Dependencies between two rules $r_1$ and $r_2$ arise when rule $r_1$ creates some node or edge which is required for the application of rule $r_2$, or when $r_1$ deletes something that is forbidden by a NAC of $r_2$. They can be defined analogously to conflict critical pairs.

CPA is essentially a brute force calculation of all critical pairs for all possible pairs of rules. From this, it is possible to present an enumeration of all potential conflicts/dependencies between rule applications or, alternatively, to display this information visually as in Figure~\ref{fig:cpa-UC1}. This is called a \emph{CPA graph}: nodes represent the rules and edges represent possible conflicts (solid lines) and dependencies (dotted lines).

\medskip

\begin{example}
Figure \ref{fig:cpa-UC1} shows the conflicts and dependencies between rules from Figure \ref{fig:gg-from-uc}.  It is possible to see that all rules are in conflict with themselves. This represents the fact that, after a match is found and the rule is applied, it cannot be applied again in the same part of the graph. In a UC, this means that each step of the UC is supposed to be performed only once. Rules \textsf{ValidateCard} and \textsf{InvalidCard} are related by an undirected conflict line: they are in \textbf{dd} conflict. In this case, this is an expected conflict, as these rules represent a decision point of the UC. Rule \textsf{InsertCard} is in \textbf{ff} conflict with itself, which is also expected as one should not insert a card after it has  already been inserted. This graph was automatically generated from the rules by the AGG tool, which can also inform which type of conflict/dependency each arrow represents and show the elements that caused the conflict/dependency. 
 
%
%
%


\begin{figure}[htbp]
\centering
\includegraphics[width=.8\textwidth]{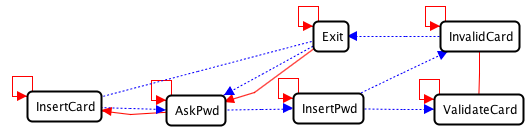}
\caption{Conflicts and Dependencies from Figure 2.}
\label{fig:cpa-UC1}
\vspace{-0.3cm}
\end{figure}
\end{example}

Each  unexpected conflict/dependency in the CPA graph represents a possible error either in the UC text or in its corresponding model. Thus, by checking each edge of the CPA graph, an analyst performs a guided verification of both the UC original text and its model. Problems can then be immediately corrected in either (or both) artifact(s). In \cite{LNWADT2015}, we provided a list of OIs that may arise from the analysis of this graph. When all issues are solved, the UC must be manually updated according to the changes made to the rules so as to improve it and keep it up-to-date.





\color{black}





\color{black}

\subsection{Second-order Graph Transformation}
\label{sec:sogt}

In this section we review the notion of rule-based modification of typed graph rules with and without NACs presented in \cite{machado2012, Machado2015}. This  rewriting of rules is based on the DPO approach, thus the rule format remains : $L \leftarrow K \rightarrow R$. However, instead of rewriting graphs, rules will be used to rewrite other rules. This new rule scheme is called second-order rule, or 2-rule for simplicity, and
it requires the definition of  morphisms between rules. 

\begin{defn}[span, span morphism] 
A (typed graph) \emph{span} is a diagram with shape $G \xleftarrow{l} G' \xrightarrow{r} G''$ in the category of $T$-typed graphs. For convenience, we refer to spans as  the pair of morphisms $(l,r)$ with common source.  A \textbf{span morphism} $f:s \to s'$ between spans $s = (l,r)$ and $s' = (l',r')$ is a triple $(f_L,f_K,f_R)$ of typed graph morphisms between the objects of the spans such that the diagram below commutes. 

\ 

\centerline{\xymatrix{
		L\ar@{}[dr]|{=}\ar[d]_{f_L} & K\ar@{}[dr]|{=}\ar[d]^{f_K}\ar@{>->}[l]_l\ar@{>->}[r]^r\ar[d] & R\ar[d]^{f_R} \\
		L' & K'\ar@{>->}[l]^{l'}\ar@{>->}[r]_{r'} & R'}}

\ 	
	
\noindent A rule morphism is mono/epi/isomorphic if all three morphisms are also mono/epi/isomorphic. Spans and span morphisms constitute a category, named \textbf{Span}. 
\end{defn} 

\begin{defn}[rule, rule morphism]
A (typed graph) \emph{rule} $r$ is a span $L \xleftarrow{l} K \xrightarrow{r} R$ such that $l$ and $r$ are \emph{monomorphisms}, i.e. injective typed graph morphisms. 
A \emph{rule morphism} $f: r \to r'$ is a span morphism $(f_L,f_K,f_R)$ between rules. Notice that $f_L$, $f_K$ and $f_R$ need not be injective. 
\end{defn}

\begin{defn}[second-order rule (2-rule)]
A second-order rule  is  span of monomorphic rule morphisms.
A second-order rule  with NACs is a pair $(s,\nac_s)$  composed by a second-order rule  $s$ and a set of (second-order) NACs for $s$, where a  second-order NAC is
 defined as a rule morphism with source on the left hand side of  $s$.

\end{defn}

Second-order rules are analogous to first-order rules, the difference is that, instead of defining transformations of graphs, they define transformations of rules (that transform graphs). Examples of second-order rules will be given in Section \ref{sec:hogt} defining evolutions the use case illustrated in Section \ref{sec:methodology}.

The transformation of a rule by means of a 2-rule is defined by means of a DPO diagram in the category \textbf{Span}, as in the case of graphs. One caveat exists, however: the resulting span may not be a valid rule because injectivity is not necessarily preserved by the rewriting. To mitigate this, it is possible to build (for each individual 2-rule) a set of \emph{structure preserving} NACs that forbids any match that would result in a ill-formed rule. In the following, we omit this set of structure-preserving NACs because they only affect the selection of 2-rule matches, not the second-order rewriting itself. The construction of this set of NACs is detailed in \cite{machado2012, Machado2015}.

\begin{defn}[second-order transformation, rule evolution]\label{def:sot}
Consider a second-order rule $\alpha = (L_{\{L,K,R\}} \leftarrow K_{\{L,K,R\}} \rightarrow R_{\{L,K,R\}})$, and a first-order rule $p = (L \leftarrow K \rightarrow R)$, as in the diagram below. The upper part of this diagram is a 2-rule
. A second-order match is a rule morphism from $L_{\{L,K,R\}}$ to $p$. Let $l = (L_L \leftarrow K_L \rightarrow R_L)$, $k = (L_K \leftarrow K_K \rightarrow R_K)$ and $r = (L_R \leftarrow K_R \rightarrow R_R)$. A second-order transformation $p \xRightarrow{\alpha,m_{\{1,2,3\}}} p''\left(L'' \leftarrow K'' \rightarrow R''\right)$ is defined by the diagram below, where $L \xRightarrow{l,m_1} L''$, $K \xRightarrow{k,m_2} K''$ and $R \xRightarrow{r,m_3} R''$ are typed graph transformations and $(L' \leftarrow K' \rightarrow R')$ and $(L'' \leftarrow K'' \rightarrow R'')$ are valid typed graph rules.

\ 

\centerline{\xymatrix@R=0.2cm@C=0.2cm{
&& L_R\ar[ddd]_{m_3} &&& K_R\ar[ddd]\ar[lll]\ar[rrr] &&& R_R\ar[ddd] \\
& L_K\ar[ddd]_{m_2}\ar[ur]\ar[dl] &&& K_K\ar[ddd]\ar[ur]\ar[lll]\ar[rrr]\ar[dl] &&& R_K\ar[ddd]\ar[ur]\ar[dl] \\
L_L\ar[ddd]_{m_1} &&& K_L\ar[ddd]\ar[lll]\ar[rrr] &&& R_L\ar[ddd] && \\
&& R &&& R'\ar[lll]\ar[rrr] &&& R'' \\
& K\ar[ur]\ar[dl] &&& K'\ar[ur]\ar[lll]\ar[rrr]\ar[dl] &&& K''\ar[ur]\ar[dl] \\
L &&& L'\ar[lll]\ar[rrr] &&& L'' && \\}}

\ 

\ 

The span of rules $p \leftarrow p' \rightarrow p''$ (the floor of the diagram above) is called \emph{rule evolution}.

\end{defn}

\section{Second-order transformations and NACs}
\label{sec:sogt-nacs}



\emph{Negative Application Conditions (NACs)} are widely used in practice, being very important in the modelling of real systems.  In particular, conditional constructs within a use case are mapped in a very direct way to rules with NACs. 

The fact that second-order rewriting as defined in \cite{machado2012} does not provide support for modifying rules containing NACs is an important limitation for its applicability.
To be able to employ second-order principles to describe and analyse evolution of use cases, our formal setting needs to support the transformation of rules with NACs. 
Formally, we need a solution for the following problem:

\paragraph{\textbf{Evolution of first-order NACs}} Given a rule with NACs $(p,NAC_p)$ and a second-order transformation $p \Rightarrow p''$ transforming $p = L \gets K \to R$ into $p'' = L'' \gets K'' \to R''$,  the question is how to obtain a set $NAC_{p''}$ to build a rule with NACs $(p'', NAC_{p''})$ maintaining the same semantics of the set  $NAC_{p}$ with respect to $p$.

~




For this analysis in particular, notice that the full second-order DPO diagram is not needed: we only require the respective rule evolution. Therefore, given a rule evolution transforming $p$ into $p''$, and a set of NACs for $p$, we intend to obtain a set of NACs over $p''$ that, ideally should forbid and allow exactly the same corresponding matches: i.e.,  the semantics of the NACs is preserved by the transformation. 
As the preconditions (the left-hand side) of the rules can be modified by a second-order rule,  the set of NACs must be altered in a compatible way.

A solution for this problem has been found recently in the context of a master thesis \cite{andrei2019}. In this (rather technical) section we describe this solution, which serves as basis for our implementation. We show that it is possible to generate a new set of NACs provided that we restrict our matches to be injective. 

We start formalizing the notion of semantic preservation, and characterize the situations in which such a semantics preservation occur. For a fixed graph $G$, we start by defining when matches  rules $p$ and  $p''$ in $G$ are related.

\begin{defn}[Related matches]
	Given a rule evolution $p \xleftarrow{l} p' \xrightarrow{r} p''$ (where $p = L \gets K \to R$, 
	$p' = L' \gets K' \to R'$ and $p'' = L'' \gets K'' \to R''$), a typed graph morphism $m$: $L \rightarrow G$ (representing a match for $p$ in $G$) and a typed graph morphism $m''$: $L'' \rightarrow G$ (representing a match for $p''$ in $G$). 
	We say that $m$ and $m''$ are \textit{related matches} (by means of the rule evolution) if and only if 
	\begin{itemize}
		\item $m$ satisfies DPO gluing conditions for $p$
		\item $m''$ satisfies DPO gluing conditions for $p''$
		\item the equation $m \circ l_L = m'' \circ r_L $ holds, i.e. the following diagram commutes 
	\[ \xymatrix@R=0.3cm{
					&& R   && R'\ar[ll]_{l_R}\ar[rr]^{r_R}  && R''  \\
			&	\ar[ur]\ar[dl]	K   && K'\ar[ll]_{l_K}\ar[rr]^{r_K} \ar[ur]\ar[dl] && K'' \ar[ur]\ar[dl] \\
					L\ar[dddrr]_{m}  && L'\ar[ll]_{l_L} \ar[rr]^{r_L}  && L''\ar[dddll]^{m''}  \\ \\ \\
					~ && G && ~ \\
				}
	\]
	\end{itemize}

\end{defn}

Intuitively, matches of a rule and its evolution are related when they are essentially the same considering the part of the left-hand side of the rule that was preserved by the rule evolution.
The following definition describes the meaning of semantics preservation of NACs in an evolution.

\begin{defn}[Preservation of NAC-behavior]
	
	Let $p \gets p' \to p''$ be a rule evolution, $NAC_p$ be a set of NACs for $p$ and $NAC_{p''}$ be a set of NACs for $p''$.

	We say that
	\begin{itemize}
		\item $NAC_{p''}$ preserves the NAC-blocking behavior of $NAC_p$ when, for every $m : L \to G$ and related $m'' : L'' \to G$, we have
		 \[  NAC_p \not \vDash m  \Rightarrow NAC_{p''} \not \vDash m'' \]
		 
		\item $NAC_{p''}$ preserves the NAC-allowing behavior of $NAC_p$ when, for every $m : L \to G$ and related $m'' : L'' \to G$, we have
		\[  NAC_p \vDash m  \Rightarrow NAC_{p''} \vDash m'' \]

		\item $NAC_{p''}$ preserves the NAC-behavior of $NAC_p$ whenever $NAC_{p''}$ preserves the NAC-blocking behavior and NAC-allowing behavior of $NAC_p$. 
		
	\end{itemize}
	
\end{defn}

The purpose of NACs is to enable or disable a match $m$ of rule $r$ in graph $G$. Roughly speaking, in DPO, a match is forbidden if (at least) one of the following situations occur:
\begin{itemize}
	\item $G$ has some element  which does not appear in (the image of)  $L$ but appear in (the image of) a NAC;
	\item  $m$ identifies   elements preserved by $r$, but these elements are not identified in a NAC;
	\item  $m$  does not identify elements  preserved by $r$, but these elements are identified in a NAC.
\end{itemize}

\begin{figure}[!ht]
	\centering
	\includegraphics[scale=.6]{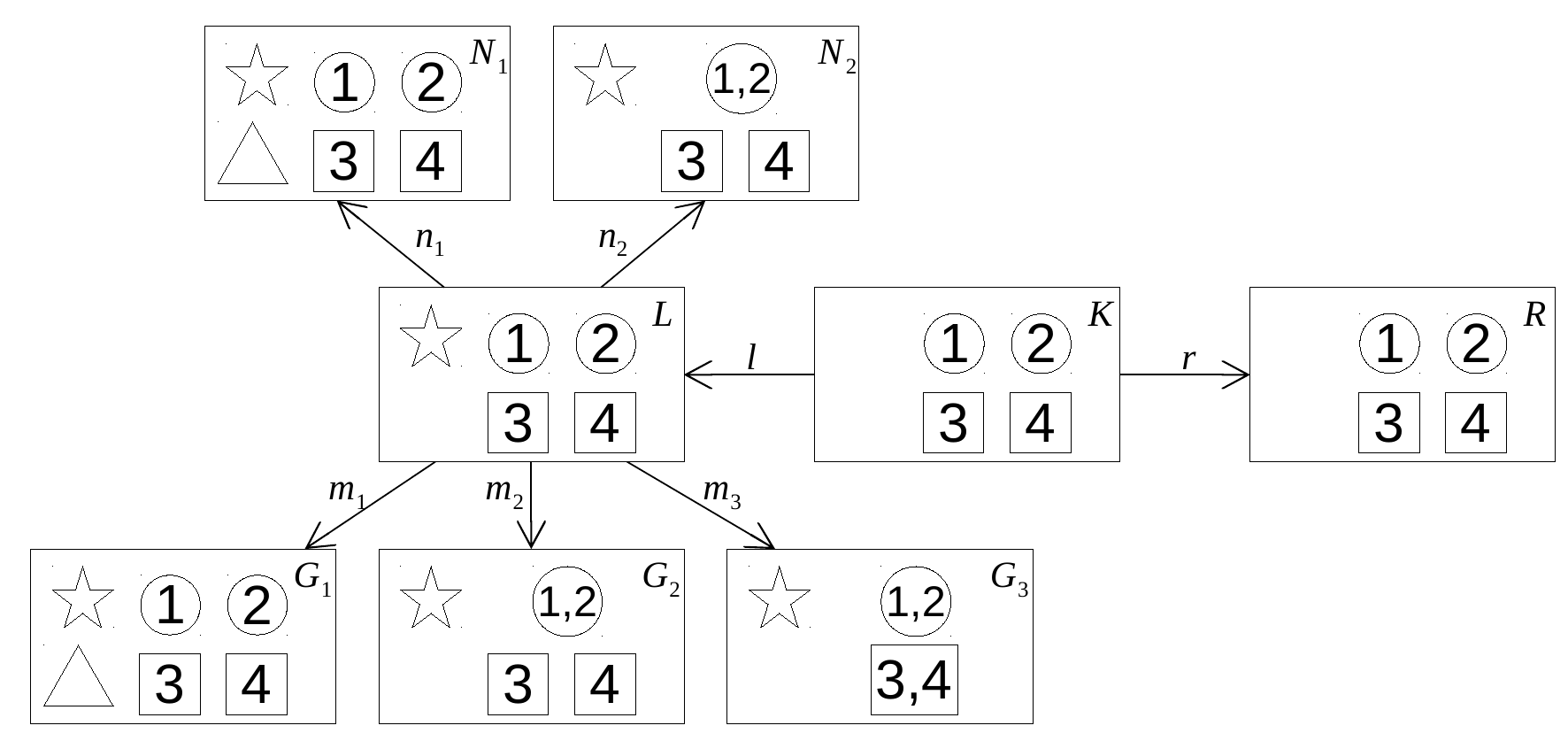}
	\caption{NACs enabling and disabling matches.}
	\label{fig:example1}
\end{figure}

These effects can be observed in the Figure~\ref{fig:example1}, which shows a rule $L \gets K \to R$ with set of NACs $\{n_1,n_2\}$ together with three graphs $G_1$, $G_2$ and $G_3$, together with one match for each graph: $m_1 : L\to G$, $m_2 : L \to G_2$, and $m_3 : L \to G_3$. The mappings are provided by the numbers within nodes. Match $m_1$ is disabled by NAC $n_1$ due to the presence of a star in $G_1$. Match $m_2$ is disabled by NAC $n_2$ due to the identification of the circles. Match $m_3$ is not disabled by $n_2$ because $n_2$ does not  identify the squares, even considering that $m_3$ identifies the circles as specified by $n_2$. 

Given a rule evolution $p \gets p' \to p''$ and a set $NAC_p$ of negative application conditions for $p$, we now define how to build a set $NAC_{p''}$ of NACs for $p''$. We will show that 
$NAC_{p''}$ preserves the NAC-blocking behavior of $NAC_{p}$. We also present a counter-example that shows that a NAC-allowing behavior preservation is not possible in general, considering arbitrary evolutions, and propose a restriction to guarantee NAC-allowing behavior preservation.

In \cite{lambers2010}, Lambers defined a way to construct, from a graph $A$ with NACs, a set of equivalent NACs for a graph $B$ related to $A$ via a morphism $t : A \rightarrow B$. This construction is called \emph{NAC-shift}.

\begin{defn}[NAC-shift (over a morphism)]

Given a NAC $n : A \rightarrow N$, a morphism $A \rightarrow B$ and the diagram below, (1) is a NAC-shift if:

(i) (1) commutes.

(ii) $t'$ and $n'$ are jointly epi.

(iii) $t'$ is mono.

\centerline{\xymatrix{
N\ar[r]^{t'} & N'\ar@{}[dl]|{(1)} \\
A\ar[r]_{t}\ar[u]^{n} & B\ar[u]_{n'} \\
}}

\end{defn}

\begin{defn}[Shift of a NAC along a morphism]

Given a NAC $n : A \rightarrow N$ and a monomorphism $t : A \rightarrow B$:

$D_t\left(n\right) = \left\{n' \mid n': B \rightarrow N', t': N \rightarrow N' \right\}$ where (1) is a NAC-shift.

\end{defn}

\begin{defn}[Shift of a set of NACs]

$D_t\left(\nac\right) = \bigcup\limits_{n \in \nac} D_t\left(n\right)$

\end{defn}

Using NAC-shift it was possible to describe  the NACs evolution process as a transformation of two steps: first a pushout complement (as in DPO transformations), and then the shift of NACs over a morphism. Note that in this process each NAC can be evolved to zero or many NACs, which is not an issue since our aim is only to preserve the behavior of the NAC, and not the NACs themselves. This construction was shown in \cite{lambers2010}  to preserve the NAC-blocking behavior.

\begin{defn}[Evolution of a set of NACs]\label{def:evolSetNacs}
		Let $p \gets p' \to p''$ be a rule evolution, where $p = L \gets K \to R$, $p' = L' \gets K' \to R'$ and $p'' = L'' \gets K'' \to R''$. Let $NAC_p$ be a set of NACs for $p$. We define 
		the \emph{evolved set of NACs} $NAC_{p''}$ as 
		 \[ NAC_{p''} =  D_t(NAC_p) \] 
\end{defn} 

\begin{theorem}[Preservation of NAC-blocking behavior]\label{the:block-preserve}
	
	Let $p \gets p' \to p''$ be a rule evolution, where $p = L \gets K \to R$, $p' = L' \gets K' \to R'$ and $p'' = L'' \gets K'' \to R''$. Let $m : L \to G$ and $m'' : L'' \to G$ be related matches. Let $NAC_p$ be a set of NACs for $p$, and $NAC_{p''} =  D_t(NAC_p)$. 
	
~	

\centering{ 
	If $\nac_{p}\not\vDash m \left(p,G\right)$ then $\nac_{p''}\not\vDash m'' \left(p'',G\right)$.
}
%

	\begin{proof}
		See ~\cite{lambers2010}.	
	\end{proof}

	\end{theorem}

However, the preservation of NAC-allowing behavior is not as straightforward as the preservation of NAC-blocking behavior. We start presenting a counterexample involving non-injective matches.

\begin{figure}[!ht]
	\centering
	\includegraphics[scale=.715]{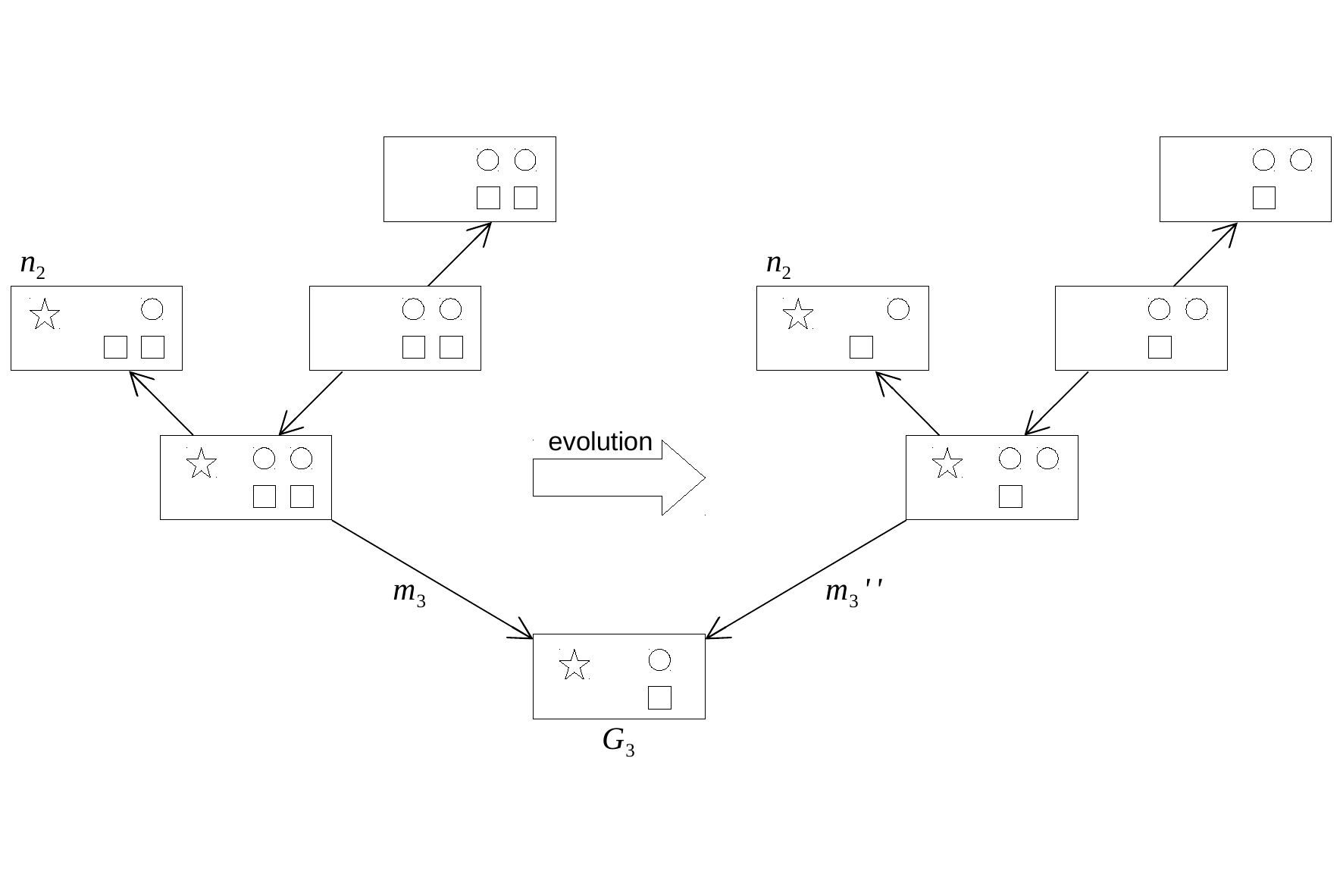}
	\vspace{-0.9cm}
	\caption{Evolution where NAC-allowing behavior is not preserved.}
	\label{fig:example2}
\end{figure}

\begin{exmp}[Invalidation of NAC-allowing behavior]
	
Figure~\ref{fig:example2} depicts an example of evolution of a rule with NACs. The  intermediate rule of the evolution is isomorphic to the right-hand side rule and it is omitted from the diagram. The original rule deletes a star in the presence of two circles and two squares. Its only NAC $n_2 : L \to N_2$ forbids the application whenever the preserved circles are identified. Notice that $m_3 : L \to G_3$ is not disabled by $n_2$, since there is only one square, and therefore there is no injective morphism $ e : N_2 \rightarrowtail G_3$. The evolution consists in removing one of the preserved squares from the rule, generating a rule that deletes a star in the presence of two circles and one square. The  NAC evolution generates a single  NAC $n_2'' : L'' \to N_2''$, without the deleted square. Notice, however, that what prevented  NAC $n_2$ from disabling $m_3$ was the impossibility of identifying the squares in a monomorphism. Since now  there is  only one square, there is actually a possible monomorphism $e'' : N_2'' \rightarrowtail G_3$ for $n_2''$, and therefore $n_2''$ disables the match $m_3''$, which is related to $m_3$ by the evolution. 
\end{exmp}

This situation occurs because some NACs forbid more than one (preserved) elements and if an evolution deletes some of these preserved elements, the distinction between matches that would be allowed or disabled by the NAC may disappear. 
If, however, we restrict second-order transformations to injective matches, these problematic situations are impossible, and it is possible to obtain preservation of NAC-allowing behavior, as we show in the following theorem.

\begin{theorem}[Preservation of NAC-allowing behavior for injective matches]\label{the:preservation}
	
	Let $p \gets p' \to p''$ be a rule evolution, where $p = L \gets K \to R$, $p' = L' \gets K' \to R'$ and $p'' = L'' \gets K'' \to R''$. Let $m : L \rightarrowtail G$ and $m'' : L'' \rightarrowtail G$ be related injective matches. Let $NAC_p$ be a set of NACs for $p$, and $NAC_{p''} =  D_t(NAC_p)$. 

~

\centering{ 
	If $\nac_{p} \vDash m \left(p,G\right)$ then $\nac_{p''} \vDash m'' \left(p'',G\right)$.
}

\begin{proof}~

\begin{itemize} 
	
	\item (by definition) \\
 	If  
  	for all $n : L \to N \in NAC_p$ there is no monomorphism $e : N \to G $  such that  $e \circ n = m$,
  	then 
  	for all $n'' : L'' \to N'' \in NAC_{p''}$  there is no monomorphism $e'' : N'' \to G $  such that  $e'' \circ n'' = m''$.

  \item (contrapositive) \\
  
   If there exists   $n'' : L'' \to N'' \in NAC_{p''}$ and monomorphism $e'' : N'' \to G $  such that  $e'' \circ n'' = m''$, then there exists $n : L \to N \in NAC_{p}$ and monomorphism $e : N \to G $ such that  $e \circ n = m$.

  \item (existence of monomorphism $e$) consider the diagram below:

\[  
\begin{tikzcd}
	&  &  \textcolor{blue}{H}  \arrow[dddrr, phantom, "\textcolor{blue}{\lrcorner}", at start]   &  & \textcolor{red}{G} \arrow[ll, blue, "v" description, tail]                                                                              &  &                                            &  &                                                                                                \\
	&  &                                                                           &  &                                                                                                                  &  &                                            &  &                                                                                                \\
	&  &                                                                           &  &                                                                                                                  &  &                                            &  &                                                                                                \\
	&  & N \arrow[drr, phantom, "\lrcorner", at start]  \arrow[rruuu, magenta, "e" description, tail] \arrow[uuu, blue, "u" description, tail] &  & N' \arrow[ll, "l_N" description, tail] \arrow[rr, "r_N" description, tail] \arrow[uuu, teal,  "e'" description, tail]   &  & N'' \arrow[lluuu, violet,  "e''" description, tail] &  &                                                                                                \\
	L \arrow[rru, "n" description, tail] \arrow[rrrruuuu, red, "m" description, tail, bend left] &  &                                                                           &  & L' \arrow[u, "n'" , tail] \arrow[llll, "l_L" description, tail] \arrow[rrrr, "r_L" description, tail] &  &                                            &  & L'' \arrow[llu, "n''" description, tail] \arrow[lllluuuu, red,  "m''" description, tail, bend right]
\end{tikzcd}
\]
  
  \begin{itemize}
  	\item assume monomorphisms $m : L \rightarrowtail G$, $m'' : L'' \rightarrowtail G$ and $e'' : N'' \rightarrowtail G$ in the diagram above. Notice that each $n'' \in NAC_{p''}$ was created from some $n \in NAC_{p}$ by means of a pushout complement and a NAC shift commutative square, as shown in the diagram;
  	\item $n'' : L'' \to N''$ is mono because $m'' = e'' \circ n''$ and $m''$ is mono. By a similar argument, note that $n'$ and $n$ are also mono.
  	\item let $e'  : N' \rightarrow G$  be the composition of monos $e'' \circ r_N$;
  	\item let $(u,v)$ be the pushout of $(e',l_N)$. Because the category of graphs is adhesive, pushouts preserve monomorphisms and, therefore, both $u : N \to H$ and $v: G \to H$ are mono;
  	\item let $e : N \to G$ be the unique arrow from pushout square $(n',l_L,L_N,n)$ towards the cospan $(m,e')$;
  	\item $e : N \to G$ is mono because $u = v \circ e$, and $u$ is mono.
  \end{itemize}	  
	
\end{itemize}

\end{proof}

\end{theorem} 

As Theorems~\ref{the:block-preserve} and~\ref{the:preservation} show, NACs preservation is constrained by the kind of morphism allowed as a rule match: 
\begin{itemize}
	\item with general matches, only preservation of blocking behavior is possible.
	\item with injective matches, preservation of allowing and blocking behavior is possible.
\end{itemize}
Considering this scenario, we use Definition~\ref{def:evolSetNacs} as appropriated algorithm for evolving a set of NACs along a rule evolution, and we must employ only injective matches in our graph grammars. This is the basis of our implementation of a second-order transformation model for first-order rules with NACs.

\section{Use Case Evolution}
\label{sec:hogt}

As software is always evolving, it is necessary to create ways to specify this evolution and analyze the impact of changes. In order to evolve use cases, a very simple idea would be to manually modify the UCs. However, modifying directly the UCs would mean to work on an inherently ambiguous and imprecise language rather than using the corresponding formal description (GT). Moreover, modifying the UC would make the GT model of the system useless, as consistency would not be maintained, unless a new translation was executed.  
Another aspect is that in large systems the effect of modifying one UC on the whole system is not always evident, since changes may propagate to other UCs due to dependencies.
On the other hand, by modifying the GT model directly, one can precisely define evolution (enabling automation) as well as get a detailed impact analysis, whilst keeping all specification artifacts coherent with the desired evolution. 

We propose to describe and analyze evolution using GT, updating the corresponding UCs only at the end of the process. Hence, we only go back to the UCs after we are certain that the evolution has the desired effect on the system. In this way, evolution leads to an updated formal model (GT) and corresponding updated UCs.

In order to represent evolution in GT, we use as  formal foundation  \emph{second-order rules} \cite{Machado2015}, which are rules that modify other rules (the first-order rules). In this work, we call  second-order rules \emph{evolution rules}, describing intended modifications to be applied to an existing GT. The main idea is to construct a rule that defines a modification pattern. This pattern denotes the items that must be changed and how they shall change for a given evolution. Then, this pattern can be compared to all rules of a GT model, called \emph{step rules} (rules describing steps of a UC). In case of a match, the rule is modified accordingly 
(an idea very similar to the use of aspects in programming \cite{Kiczales1997}). 
This way of formalizing the definition of an evolution guarantees a precise description, using a language similar to the one used to describe the system itself. It also enables the possibility of analysis of changes due to the evolution process (side-effects).

The proposed method  consists of 3 steps:

\noindent
\begin{description}
\item [Step 1 -- Definition of evolution:]  
      The developer defines any behavior that has to be modified, added to, or suppressed from the system in terms of GT rules. Behaviors to be modified are described as evolution rules (second-order rules), whereas new behavior is described as step rules (first-order rules). Rules to be deleted may be specified explicitly or be selected in Step 3.2.
\item [Step 2 -- Analysis of evolution:]  
      Using evolution rules, analysis based on critical pairs (CP) can be performed:
      \begin{description}
      \item [Step 2.1 -- CPs between evolution rules:]
       give hints on consistency and termination of the evolution process;
       \item [Step 2.2 -- CPs between evolution and execution:]  describe how system's step rules (and corresponding UC steps)  will be impacted by  evolution rules. By analyzing these CPs, the developer has to decide whether the evolution should be performed as is or modified to prevent undesired behavior (the latter case means going back to Step 1);   
       \end{description}
\item [Step 3 -- Execution of evolution:] 
      Perform the intended evolution:
      \begin{description}
      \item [Step 3.1 -- Generate a new GT:]
	         Execute the modifications defined in Step 1 on the original GT;
      \item [Step 3.2 -- Analyze the new GT:]
             Construct the CPA graph for the new GT and identify rules that shall be deleted (became useless), perform deletion, and redo analysis until the GT exhibits the desired behavior;
      \item [Step 3.3 -- Modify UCs:]
            UCs whose step rules were affected by evolution should be rewritten accordingly.
      \end{description}
\end{description}

Before detailing how evolution is defined and performed, we illustrate the proposed method with a simple example of evolution applied to our running. This example considers only one UC, but more complex case studies are reported in Section~\ref{sec:experiments}.

\begin{figure}[htbp]
\centering
\includegraphics[scale=0.4]{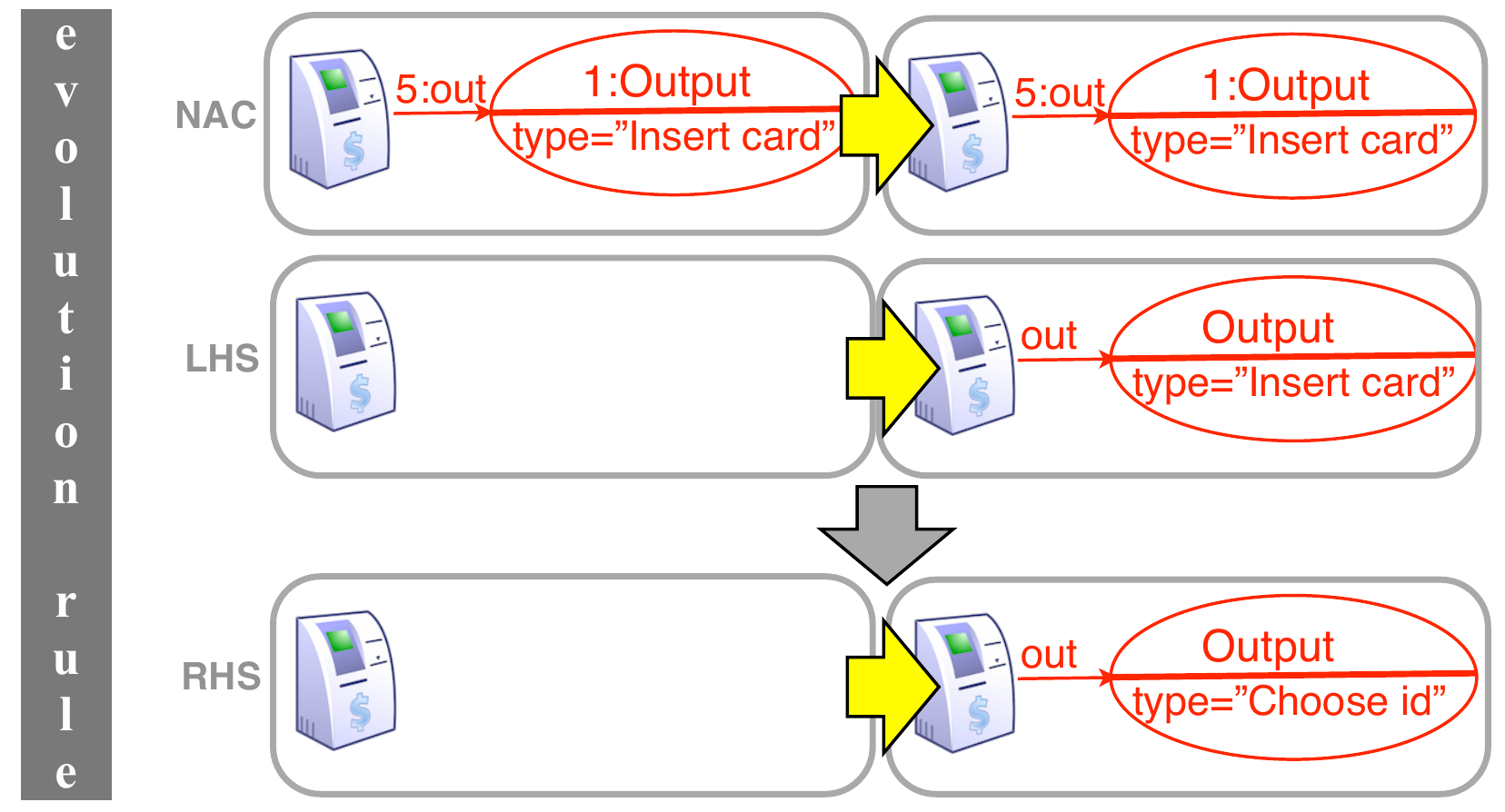}

\caption{Evolution Rule}
\label{fig:evol1}
\vspace{-0.3cm}
\end{figure}

\begin{example}(Evolution 1: Insert fingerprint login.)
As an example of evolution, suppose the bank now decides to change the way in which a client logs in via an ATM: besides the original card-and-password method, the bank also wants to allow identification via fingerprint. With this change, any step of a UC leading back to the start state of the ATM, showing the \textsf{Insert card} message, would  have to be modified to show a \textsf{Choose id} message, indicating that the user now has to select the identification method to be used.

\medskip

\noindent\textbf{Step 1:}
Figure~\ref{fig:evol1} describes an evolution rule modeling this modification. To stress the orthogonal nature of evolution with respect to the behaviour of the system, evolution rules will be depicted vertically (contrasting to the horizontal description of step rules). The LHS of the evolution rule (shown above the downward arrow) is the pattern to be found in a step rule of the specification to trigger a modification on that rule. In this example, it matches all step rules where an ATM is preserved and an \textsf{Output} node of type \textsf{"Insert Card"} is generated. The effect, as specified by the evolution rule's RHS (rule pattern below the downward arrow), is to delete the \textsf{Output} node of type \textsf{"Insert card"} and replace it by an \textsf{Output} node of type \textsf{"Choose id"}. The NAC of the evolution rule (shown on the top) is used to ensure that this transformation is applied only to step rules that create an \textsf{Output} node \textsf{"Insert Card"}, not to rules that preserve it. Besides changing existing step rules, new rules are added to model the new behavior (see Figure \ref{fig:evol1Rules}): now the system starts by asking the user to choose between card or fingerprint identification (rules \textsf{ChooseId-card} and \textsf{ChooseId-fprt}), and rules that treat insertion and validation of fingerprints are added.

\begin{figure}[htbp]
\centering
\includegraphics[scale=0.3]{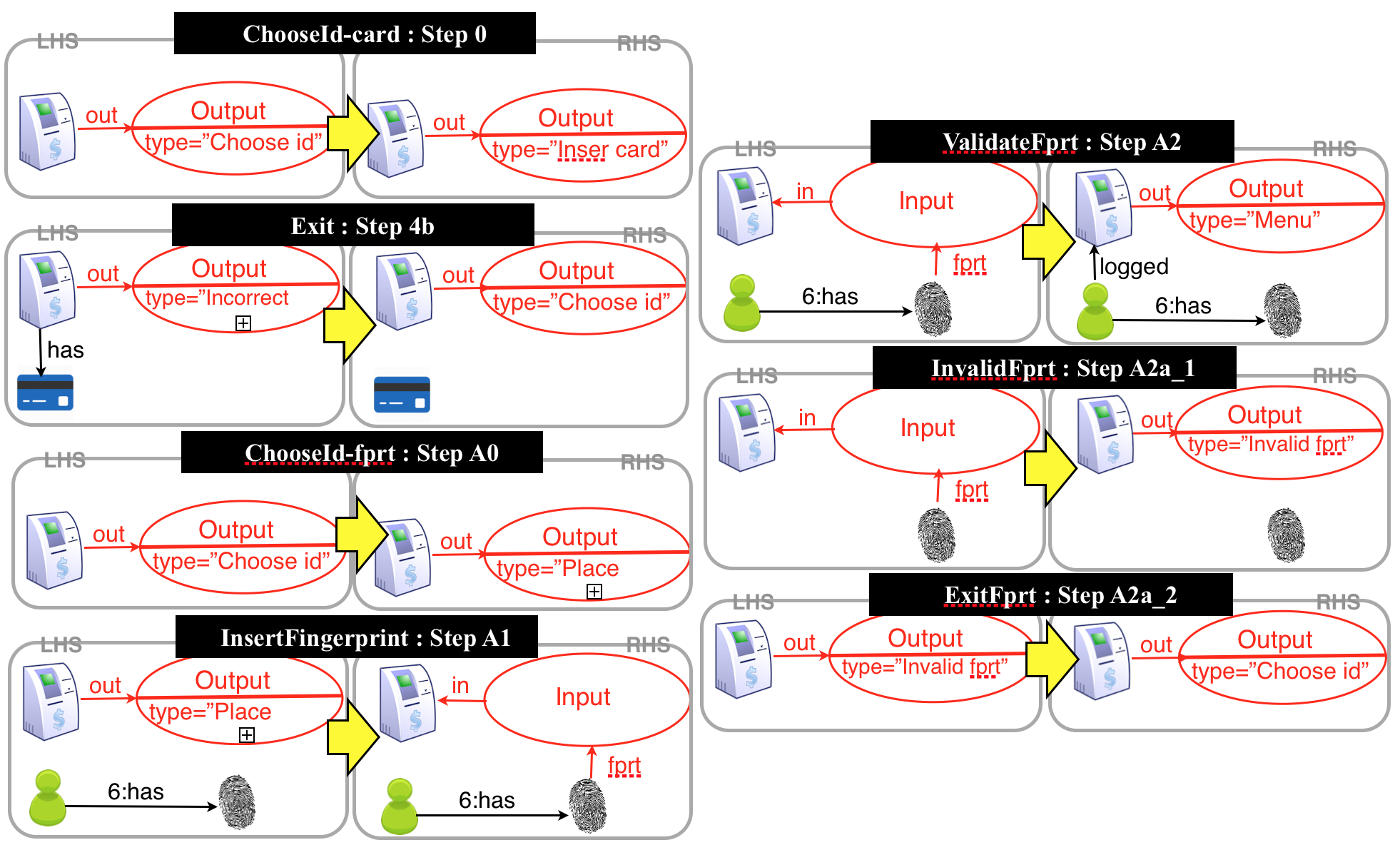}
\caption{UC Rules after application of the Evolution Rule}
\label{fig:evol1Rules}
\vspace{-0.3cm}
\end{figure}

\medskip

\noindent\textbf{Step 2:}
This evolution consists of only one rule, which is in conflict with itself because, once applied, it will change the step rule in a way that the evolution rule can not be applied again (to this step rule). Thus, the evolution process is consistent and terminates.
By analyzing the matches and conflicts of the evolution rule with the execution ones, 
it is possible to detect all step rules that should be modified and, consequently, all  UC steps that should be changed to apply the evolution. In our example, the only step rule that would be modified is rule \textsf{Exit} (corresponds to Step 4a\_2), that will now generate a message \textsf{Choose id} instead of \textsf{Insert card}. This is exactly what was expected, hence the evolution can be performed.

\medskip

\noindent\textbf{Step 3:}
We can now perform the evolution (Step 3.1) and generate the CPA graph (Step 3.2 -- see Figure \ref{fig:CPAlogin_evolution1}). In
this example no step rule shall be deleted (the next example will illustrate this case) and we can thus update the corresponding UC: steps of the original UC corresponding to unchanged step rules remain unchanged; steps that correspond to changed rules must be updated; and steps corresponding to new rules must be added. This UC is shown in Figure \ref{fig:UCLogin_evolution1}.

\begin{figure}[hbp]
\centering
\includegraphics[scale=0.4]{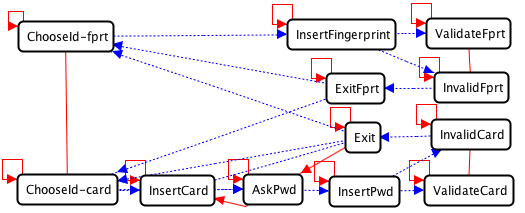}
\caption{CPA graph - evolution 1.}
\label{fig:CPAlogin_evolution1}
\end{figure}

\begin{figure}[htp]
\centering
\includegraphics[width=0.8\textwidth]{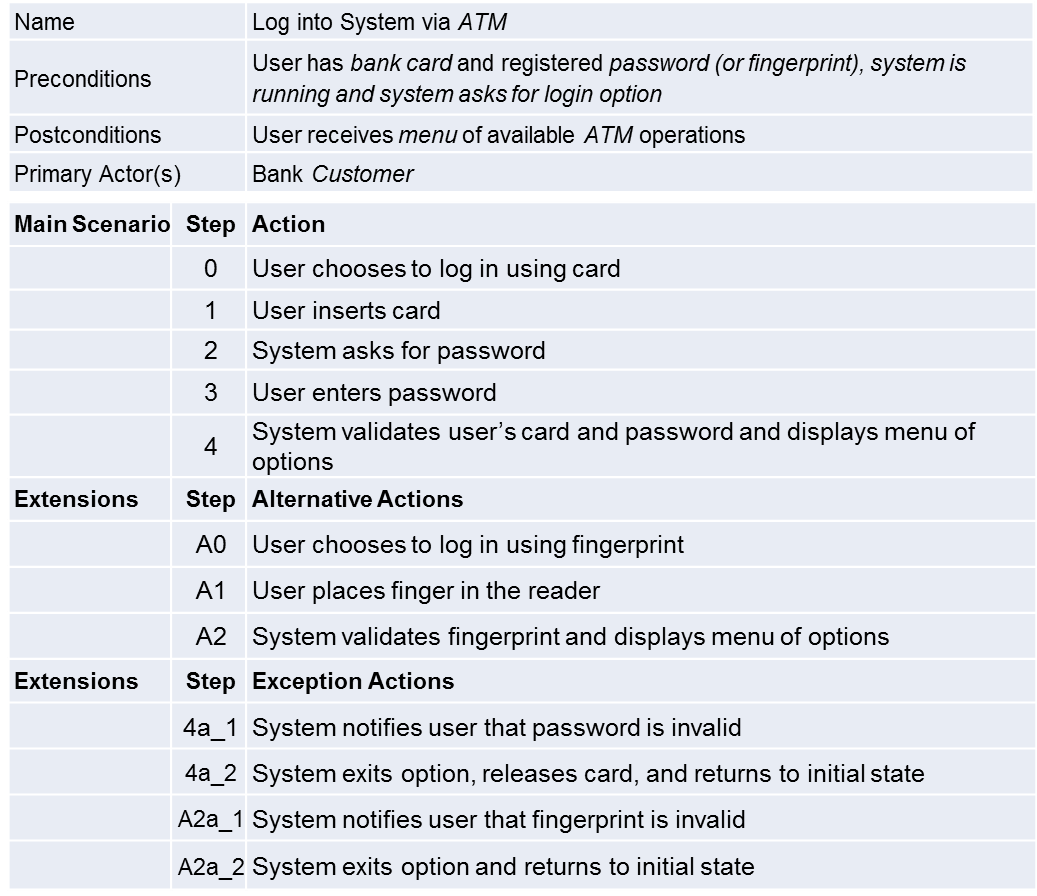}
\caption{Login Use Case description - evolution 1.}
\label{fig:UCLogin_evolution1}
\vspace{-0.3cm}
\end{figure}

\end{example}



\subsection{Formalization of Evolution}

This section presents the formalization of the evolution steps previously discussed and applied to to our ATM example.

\medskip

\noindent
\textbf{Step 1:} We now define how to represent the evolution of a system. 



When modifying a graph transformation system, 
one may alter the type graph (inserting or deleting types of elements), add new rules, delete deprecated rules and update  rules by means of second-order rules (which in this context will also be referred as evolution rules). All these modifications are codified by the structure we call \textit{evolution structure}.

\begin{defn}[Evolution structure]
\label{d.evolutionstructure}
Let $\mc{G} = (T,P)$ be a GT system. An \textbf{evolution structure for $\mc{G}$} is a tuple $ES = (ET,EP,Del,New)$ where
\begin{enumerate}
\item $ET = T \xleftarrow{l} T' \xrightarrow{r} T''$ is a span of graph morphisms such that $l$ and $r$ are injective. This describes how the type graph is modified by the evolution.
\item $EP$ is a set of second-order rules (evolution rules) that describe how to update the rules which are preserved during the evolution; the LHS of each evolution rule must be typed over $T$ and
the RHS must be typed over $T''$;
\item $Del \subseteq P$ is a set of GT rules  to be deleted;
\item $New$ is a set of GT rules ($T''$-typed) to be created.
\end{enumerate}
\end{defn}

\begin{figure}[!htbp]

\[
\xymatrix@=1cm{ 
\textbf{(a)} &   &  *++{L_{Ev}}   \ar@{->}[d]_{m_{Ev}} \ar@[white]@{}[rd]|{(1)}  
    & *++{K_{Ev}}  \ar@{}[dr]|(.8){\ulcorner} \ar@{}[dl]|(.8){\urcorner} \ar@{>->}[l]_{l_{Ev}} \ar@{>->}[r]^{r_{Ev}} \ar[d]^{} \ar@[white]@{}[rd]|{(2)}  
    & *++{R_{Ev}} \ar[d]^{}
\\  
\ & &   *++{p}
    & *++{p'} \ar[l]^{} \ar[r]_{}
    & *++{p''} & \ \qquad \ \qquad  \\
} \vspace{-0.5cm}
\]  

\[  \vspace{-0.3cm}
\xymatrix@=0.6cm{ 
\textbf{(b)}    &  *++{LHS(L_{Ev})}  \ar@{->}[d]_{m} \ar@[white]@{}[rd]|{(1)}  
    & *++{LHS(K_{Ev})}  \ar@{}[dr]|(.8){\ulcorner} \ar@{}[dl]|(.8){\urcorner} \ar@{>->}[l]_{l} \ar@{>->}[r]^{r} \ar[d]^{k} \ar@[white]@{}[rd]|{(2)}  
    & *++{LHS(R_{Ev})} \ar[d]^{m^*}
\\  
X
  &   *++{L_p} \ar@{->}[l]_{n}
    & *++{L_{p'}} \ar[l]^{l^*} \ar[r]_{r^*}
    & *++{L_{p''}} & \ \\
}
\]
\caption{(a) Span rewriting  (in \Span) (b) Rewriting of a rule's LHS (in \Typed{T}) }
\label{diagHOR}
\vspace{-0.3cm}
\end{figure}


\medskip

\begin{figure}[htbp]
\centering
\includegraphics[scale=0.4]{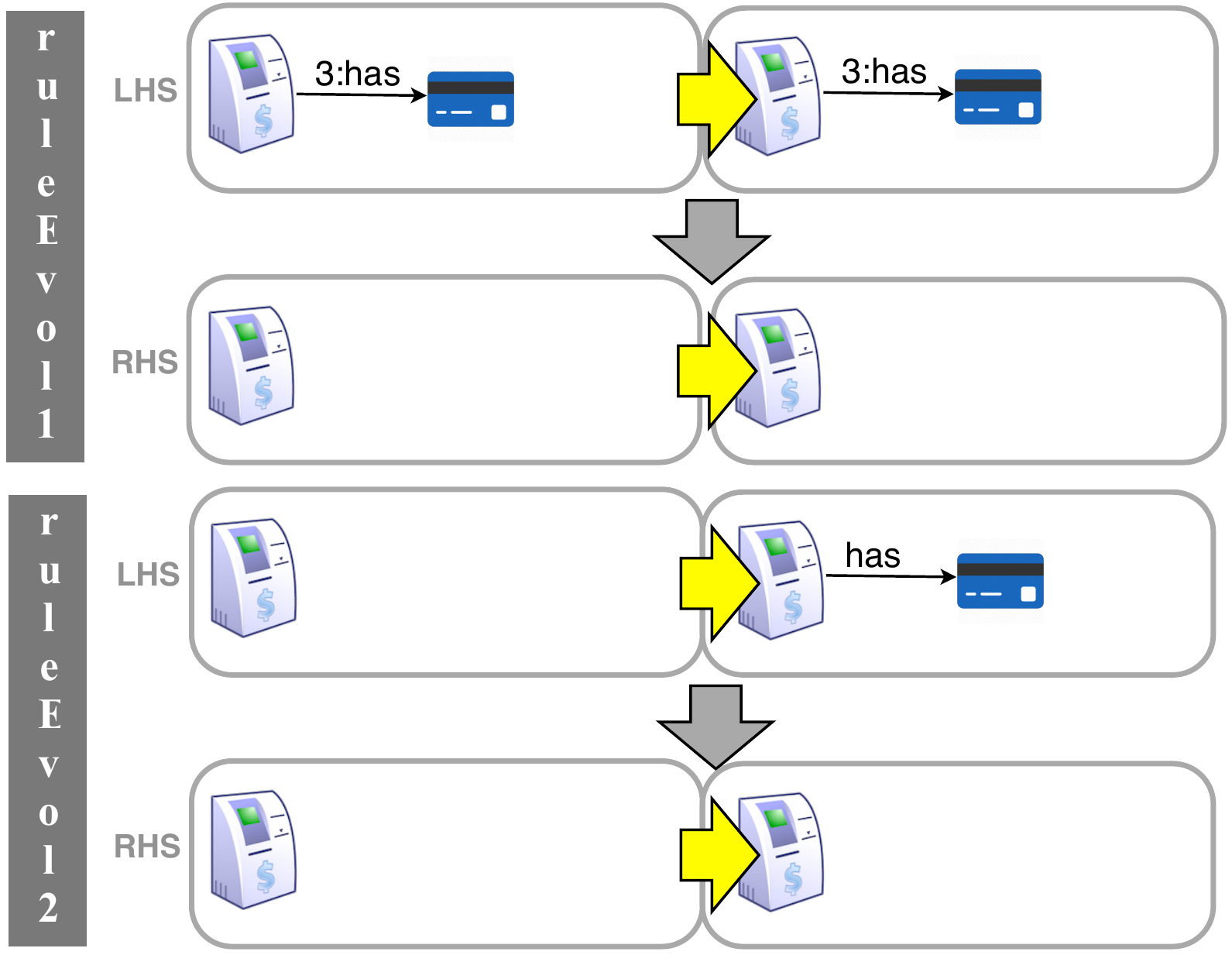}
\caption{Evolution of rules: Removing card}
\label{fig:evol2}
\end{figure}

\begin{example}
(Evolution 2: Remove card/password login) 
Suppose now that the bank decides to remove the old card-and-password method and make the fingerprint recognition the only available login method. In this case, all step rules that contain any reference to card or password would have to be modified. Figure~\ref{fig:evol2} shows examples of two evolution rules that may be used to accomplish this task. Rules \textsf{rule-Evol1} and \textsf{rule-Evol2} state that step rules that preserve/produce bank card should have bank cards removed. Similar rules can describe that references to bank cards, as well as references to \textsf{Insert card}, should be removed. We will call these rules 
\textsf{rule-Evol3} (rules that create references to cards should be updated), 
\textsf{rule-Evol4} (rules that preserve \textsf{Insert card} should be updated), 
\textsf{rule-Evol5} (rules that create \textsf{Insert card} should be updated), 
\textsf{rule-Evol6} (rules that delete \textsf{Insert card} should be updated).
We also have \textsf{rule-Evol7}, analogous to the rule shown in Figure~\ref{fig:evol1}, that rewrites step rules that create \textsf{Choose id} to rules that create \textsf{Place finger} (because now there is no choice of identification method). Finally, \textsf{rule-Evol8} removes references to \textsf{Choose id} from the LHSs of step rules. The tuple $(T\gets T' \to T', \{ \textsf{rule-Evo1},\dots, \textsf{rule-Evo8}\},\{\},\{\}\})$ is an evolution structure, where $T'$ is obtained from $T$ by removing the node types \textsf{card}, \textsf{Insert card} and associated edge types.
\end{example}

\newcommand{\rewSOR}[3]{\ensuremath{{#1}\stackrel{{#2}}{\Longrightarrow_2}{#3}}}

\medskip


\noindent\textbf{Step 2.1:} In \cite{Machado2015}, critical pairs between higher-order rules were defined. Here we can apply this notion to evolution rules obtaining a CPA graph describing the conflicts and dependencies. We require that every evolution rule be in conflict with itself and do not depend on itself. Moreover, we require that there are no conflicts nor dependencies between two different evolution rules. 
This guarantees that every rule can be applied at most once at each match and, therefore, the evolution process will eventually terminate (all rule sets and graphs are finite) and that the evolution process is deterministic.


\smallskip

\noindent\textbf{Step 2.2:} There are different ways in which an evolution rule may impact a step rule. Intuitively, the \textbf{gluing problem}  occurs when the evolution rule cannot be applied because the (second-order) gluing condition is not satisfied. This means that the application of this rule would lead to side effects; for example, when we are trying to remove some node of the step rule that is referenced by some edge not specified for deletion in the evolution rule.
The \textbf{NAC-gluing problem} represents a situation in which applying an evolution rule implies the deletion of a NAC of the step rule: if the elements that were forbidden by the NAC are removed, the NAC becomes useless.
These situations actually prevent the application of the evolution rule to this step rule since they would lead to side effects that were not foreseen. When this occurs, the developer has to decide whether the side effect is desired (in which case a new rule explicitly stating this not as a side effect but as the effect of the rule must be added) or not (in this case, nothing must be done, since the evolution rule is not applicable to the step rule in which side effects would occur).
Other situations  to consider are when the evolution rule increases or restricts the step rule applicability (the LHS of the step rule is changed), and when the evolution rule changes the effect of the step rule (the RHS of the step rule is changed). These situations are characterized below.

\begin{defn}[Interplay between Evolution and Execution rules]
\label{def:rr-avoiding-ri}
Given an evolution rule with NACs $EvR = (EvNAC,\alpha)$ that is not an isomorphism, where \mbox{$\alpha = L_{Ev} \leftarrowtail K_{Ev} \rightarrowtail R_{Ev}$}, a step rule with NACs $ExR=(exNAC,p)$ and a morphism
 $m : L_{Ev} \to p$, \textbf{the  interplay between the evolution rule $EvR$ and the step rule $ExR$} can be classified in 4 groups:
 \begin{description}
 \item [1. Situations that prevent evolution:]\ 
 \begin{description}
 \item [Gluing problem (gp):] there is no unique $p'$ such that diagram (1) of Figure \ref{diagHOR}(a) is a pushout
        \footnote{The name \emph{gluing} is standard in the graph transformation community, and denotes a situation in
        which the gluing rule $p'$ can not be determined. This may occur due to a conflict between deletion and preservation
        of items during rule application, or due to trying to delete items that are connected to others by edges that are not
        specified for deletion by the rule.};
 \item [NAC-gluing problem (np):] there is a NAC $n:L_p\to X \in N$ such that $n \circ l^*$ in Figure \ref{diagHOR}(b) is an isomorphism;
 \end{description}
 \item [2. Situations that increase rule applicability:]\ 
 \begin{description}
 \item [Delete$^2$-delete (d$^2$d):] an item deleted by $ExR$ is deleted from $p$ by $EvR$; 
 \item [Delete$^2$-preserve (d$^2$p):] an item  preserved by $ExR$ is deleted from $p$ by $EvR$;
 \end{description}
 \item [3. Situations that restrict rule applicability:]\ 
 \begin{description}
 \item [Create$^2$-delete (c$^2$d):] $EvR$ creates an item to be deleted by $p$ (i.e., inserts an item in $p$'s LHS);
 \item [Create$^2$-preserve (c$^2$p):] $EvR$ creates an item to be preserved by $p$ (i.e., inserts an item in $p$'s LHS);
 \end{description}
 \item [4. Situations that modify rule effect:]\
 \begin{description}
 \item [Create$^2$-create  (c$^2$c):] $EvR$ creates an item to be created by $p$ (i.e., inserts an items in $p$'s RHS);
 \item [Delete$^2$-create (d$^2$c):] $EvR$ deletes an item created by $p$.
 \end{description}
 \end{description}
 \end{defn}

The open issues that may be raised by this analysis step are listed in Table \ref{fig:OIEvolution}.

\begin{table*}[htbp]
\centering
{\footnotesize 
\begin{tabular}{|p{0.9cm}|p{2.5cm}|p{2.5cm}|p{1.5cm}|p{3.5cm}|} \hline
 {\bf Open issue} & {\bf Verification} & {\bf Problem} & {\bf Severity level} & {\bf Possible action}	\\ \hline
OI.Ev1 & Evolution rule not in conflict with itself 
            & The rule may be applied multiple times
            &  \vspace{0.3cm}\textbf{\includegraphics[width=3mm]{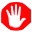} Red} 
            & Review the evolution rule (adding a NAC or an item to be deleted).\\ \hline
 OI.Ev2 & Evolution rule dependent on any evolution rule
  	    & Evolution may not terminate
	    &  \vspace{0.3cm}\textbf{\includegraphics[width=3mm]{figs/oired.png} Red} 
	    & Review the rules. \\ \hline
OI.Ev3 & Evolution rule in conflict with other evolution rule 
            & May lead to non-deterministic evolution 
            & \vspace{0.3cm}\textbf{\includegraphics[width=3mm]{figs/oired.png} Red} 
            & Check LHSs of conflicting evolution rules to make sure that they are not applicable to the same step rule\\ \hline
OI.Ev4 & Gluing problem occured
            &  An evolution is not applicable to some step rule due to extra context
            & \vspace{0.3cm}\textbf{\includegraphics[width=3mm]{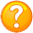} Orange} 
            &  Check whether the evolution should be applicable to this step rule. If not, nothing has to be done. If yes, create a new evolution
            rule with the extra context, if such rule does not exist.\\ \hline
OI.Ev5 & NAC-gluing problem occured
            &  An evolution would make the NAC of a step rule useless
            & \vspace{0.3cm}\textbf{\includegraphics[width=3mm]{figs/oiorange.png} Orange} 
            & Check the NAC: if it makes sense that it become useless, delete this NAC. Otherwise, review the evolution rule.\\ \hline
OI.Ev6 &  Undesired increase of (step) rule applicability
            & Delete$^2$-delete or  Delete$^2$-preserve situation occurred
            & \vspace{0.3cm}\textbf{\includegraphics[width=3mm]{figs/oiorange.png} Orange} 
            & Check the items that will be deleted from the step rule by the evolution rule  (items deleted in LHS($L_{Ev}$). \\ \hline
OI.Ev7 &  Undesired restriction of (step) rule applicability
            & Create$^2$-delete or  Create$^2$-preserve situation occurred
            & \vspace{0.3cm}\textbf{\includegraphics[width=3mm]{figs/oiorange.png} Orange} 
            & Check the items that will be inserted in the LHS of the step rule by the evolution rule  (items created in LHS($R_{Ev}$) . \\ \hline        
OI.Ev8 &  Undesired modification of (step) rule behavior
            & Create$^2$-create or  Delete$^2$-create situation occurred
            & \vspace{0.3cm}\textbf{\includegraphics[width=3mm]{figs/oiorange.png} Orange} 
            & Check the items that are created in/deleted from the RHS of the step rule by the evolution rule  (items created in/deleted from  RHS($R_{Ev}$). \\ \hline   
\end{tabular}}
\caption{Open Issues concerning Evolution \label{fig:OIEvolution}}
\end{table*}

\medskip

\begin{example}
Many step rules in the grammar of Figure \ref{fig:evol1Rules} will be affected by evolution 2 (removal of card and password authentication). Due to space limitations, it is not possible to show all cases, but  only examples of  situations that arise from this evolution:
\begin{description}
\item [Gluing problem:] There is a gluing problem between rule \textsf{ruleEvol2} and rule \textsf{insertCard}:  when we try to delete the card from the step rule \textsf{insertCard}, the  gluing problem warns that there is a connection of this card to a User and thus, by removing the card, we would have to remove this connection as a side effect. 
If  this is the desired effect, either this evolution rule should be updated to include the card, or another
evolution rule including the card should be added (in case the two methods -- with an without card -- are needed). Another possibility would be to mark the step rule to be deleted by the evolution (in this example, this would be the desired effect since
without the card the rule do nothing, as shown in Figure \ref{fig:evolinsertcard}).
\item [NAC-gluing problem:] Once again, the application of rule \textsf{ruleEvol2} would impact in rule \textsf{insertCard}. The NAC would become equal to the LHS of the rule, meaning that it would not forbid anything, becoming useless.
\item [Delete$^2$-delete:] rule \textsf{AskPwd} is in \textbf{d}$^2$\textbf{d}-situation with \textsf{rule-Evol6} because both delete the node \textsf{Output:InsertCard}.
\item [delete$^2$-preserve:] rule \textsf{rule-Evol4} is in d$^2$p-situation with rule \textsf{InsertCard}.
\end{description}

\begin{figure}[htbp]
\centering
\includegraphics[width=.7\textwidth]{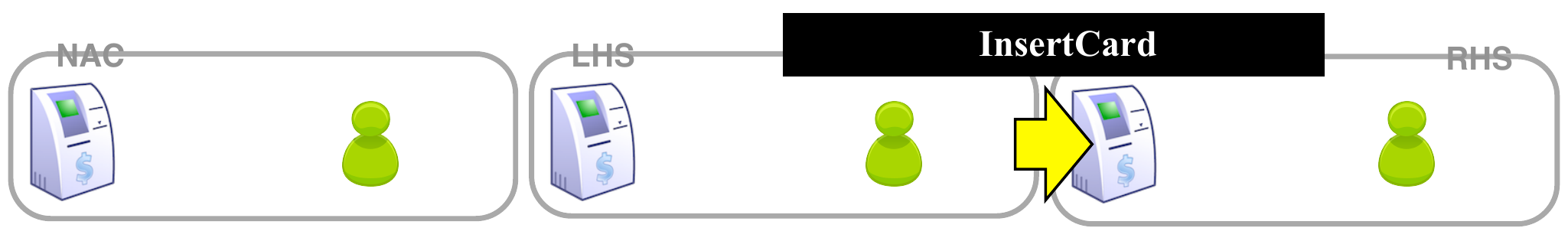}
\caption{Evolution of rule \textsf{insertCard}}
\label{fig:evolinsertcard}
\end{figure}

The existence of such conflicts means that the evolution step will change the behavior of the affected rules and, therefore, the developer has to check whether the outcome of the evolution captures the desired behavior. If, despite the conflicts, evolution according to the given rules is performed, this means the developer judged that  these conflicts do not represent undesired UC behavior.  
\end{example}

\medskip

\noindent\textbf{Step 3.1:} Let $\alpha = L_{Ev} \gets K_{Ev} \to R_{Ev}$ be an evolution rule, let $p$ be a graph transformation rule without NACs, and let $m_{Ev}: L_{Ev} \to p$ be a span morphism. The evolution of $p$ to $p''$ via the application of $(\alpha,m_{Env})$,  denoted by $p \xRightarrow{\alpha,m_{Env}} p''$, was defined in \cite{Machado2015}. 
We  now extend this notion of evolution to step rules with NACs. If $p \xRightarrow{\alpha,m_{Env}} p''$, then $(N,p) \xRightarrow{\alpha,m_{Env}} (N',p'')$, where $N'$ is obtained from $N$ as follows: for each NAC $n:L_p \to X \in N$, we can obtain a corresponding NAC $n': L_{p''}\to X'$ as the universal pushout morphism from pushout (2) in Figure   \ref{diagHOR}(b), 
where $X'$ is the pushout (object) from   $n\circ l^* \circ k$ and $r$. With this, we can define how an evolution structure evolves a GT system as a whole.
\begin{defn}[Evolution]
Let $\mc{G}$ be a graph transformation system, and $ES = (T\gets T' \to T'',EP,Del,New)$ be an evolution structure over $\mc{G}$. An \textbf{evolution} of $\mc{G}$ induced by $ES$ is the GTS $G'' = (T'',P'')$ where $P''$ is obtained by
\begin{enumerate}
\item $P_1 = P \setminus Del$, i.e., remove all step rules marked for deletion;
\item Keep applying  evolution rules in $EP$ to  step rules in $P_1$, until no evolution rule is applicable. This gives rise to the set of rules $P_2$;
\item $P'' = P_2 \cup New$, i.e., add new rules to the specification.
\end{enumerate}
\end{defn}

\smallskip
 
 \noindent\textbf{Step 3.2:} This step is a garbage collection step that deletes the step rules that are no longer useful in the system. It is performed by constructing the CPA graph of the result of the evolution and comparing it to the CPA graph of the system before evolution. Rules that are disconnected in this graph are the first candidates for deletion (since they do not use/create any item that is needed by other rules and thus represent actions that are totally independent from others, which is not common in UCs). The second verification concerns branching points: if a branching point (bi-directional conflict) disappeared, probably all rules depending on one of the branches should be removed (because UCs do not have usually two independent execution threads). Third, if a dependency disappears, it should be checked whether the rule is still necessary. Note that selecting the rules to be deleted is a manual task, since it involves choosing which behavior should be kept in the UC. However, the CPA graph helps finding the rules to be deleted. Besides using the CPA graph, we also look for rules that turned into isomorphisms due to evolution (and thus, became useless) and also for rules that are duplicated.

\smallskip

 \noindent\textbf{Step 3.3:} This step is done by removing from the original UC all steps that correspond to deleted rules, modifying the steps that correspond to changed rules (by comparing the old and new rules, it is usually obvious how the text should be rewritten), and add steps corresponding to new rules. The order of steps must be compatible with the dependency order of the CPA graph of the evolved system.

\medskip

\begin{example}(Evolution 2: Remove card/password login -- Step 3)
By performing an evolution using evolution rules \textsf{rule-Evol1}  to \textsf{rule-Evol8} and deleting step rule \textsf{InsertCard}, we obtain the modified step rules shown in Figure \ref{fig:evol2rr} (rules corresponding to other steps remain unchanged). After the application of evolution, the analyst may review the specification in order to check whether the remaining rules make sense in the new context. 
\begin{figure}[!h]
\centering
\includegraphics[width=\textwidth]{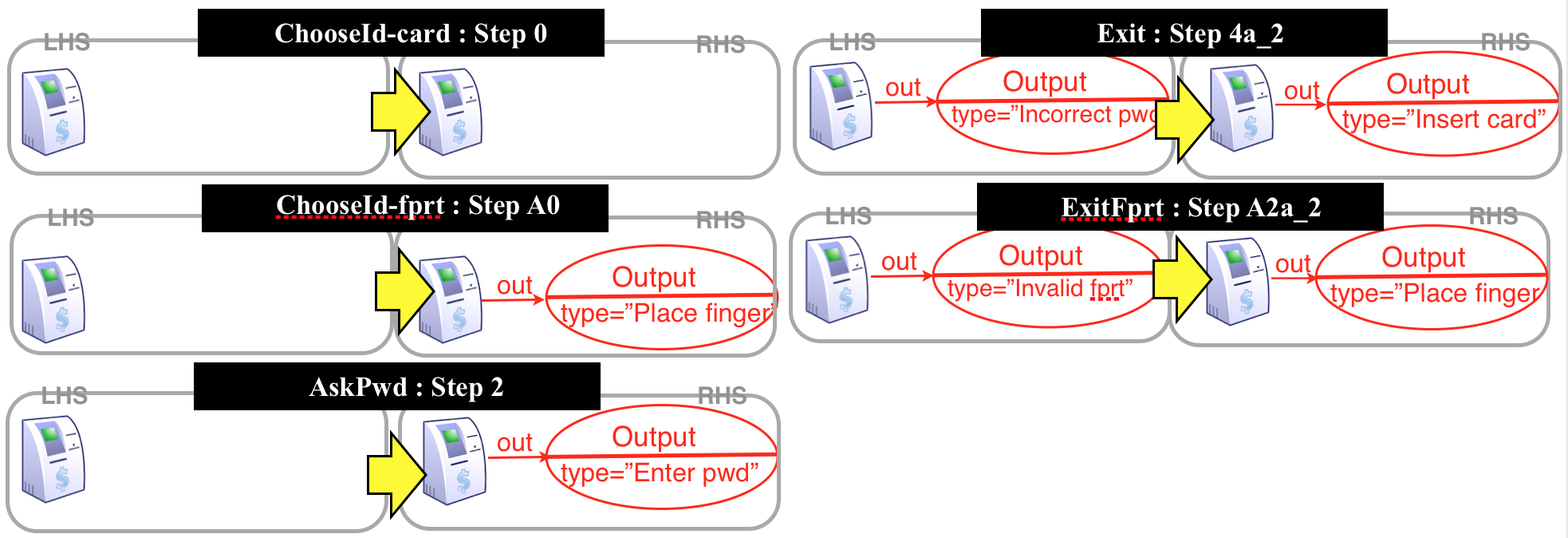}
\caption{Evolution of rules of UC1 }
\label{fig:evol2rr}
\end{figure}

The CPA graph of the resulting GT is shown in Figure \ref{fig:cpaevol2} (we
included \textsf{InsertCard} here to show how it would appear, if it had not been explicitly deleted by the evolution step).
Rules \textsf{InsertCard} and \textsf{ChooseId-card} should be removed from the set of resulting step rules because they have no effect (they do not delete/create anything, and are thus disconnected in the CPA graph). Rule \textsf{ChooseId-fprt} is not in conflict with itself and thus can occur indefinitely many times: since this choice is actually no longer needed, it may be removed. The starting point of the corresponding UC should be \textsf{InsertFingerprint}.
Rules \textsf{AskPwd}, \textsf{InsertPwd}, \textsf{ValidateCard}, \textsf{InvalidCard} and \textsf{Exit} are not reachable from \textsf{InsertFingerprint}. Since these actions are not to be performed in parallel with others (if this would be the case, being independent from \textsf{InsertFingerprint} would be correct), all these rules  should also be removed. The resulting UC is shown in Figure \ref{fig:UCLogin_evolution2}.

\begin{figure}[!h]
\centering
\includegraphics[scale=0.4]{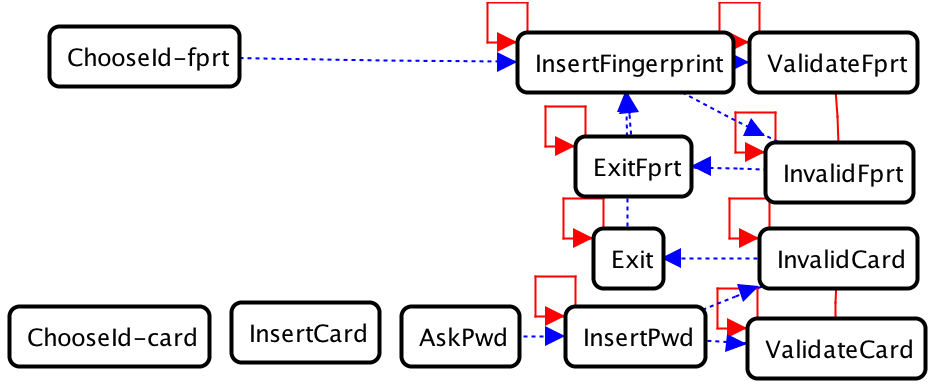}
\caption{Conflict-dependency graph - evolution 2.}
\label{fig:cpaevol2}
\end{figure}

\begin{figure}[!h]
\centering
\includegraphics[width=.8\textwidth]{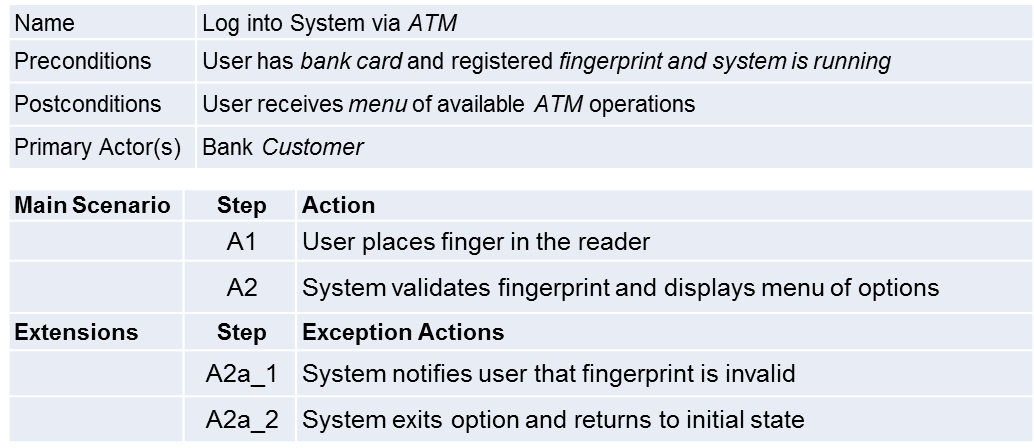}
\caption{Login Use Case description - evolution 2.}
\label{fig:UCLogin_evolution2}
\end{figure}
\end{example}

\color{black}

\section{Case Studies} \label{sec:experiments}
We present  3 case studies to evaluate the use of our approach. 
For each case study, we have previously constructed the original GT using the methodology described in \cite{LNWADT2015}. Hence, here we only describe the evolution process of these systems, as well as quantify the impact of the modifications. All  original UCs and GTs, the evolution rules, and the modified GTs can be found at \url{http://www.ufrgs.br/verites}. We used the Verigraph tool\footnote{Available at \url{http://github.com/verites/verigraph}} to perform analyses on each UC. Verigraph is implemented in the Haskell programming language and offers (amongst other functionalities) critical-pair analysis and evolution of graph transformation systems based on second-order rules. It is currently under development and, although freely available, provides at the moment only a command-line interface. We used the AGG tool \cite{Taentzer2000} to visualise analysis results and to model first- and second-order graph transformation rules.



The first case study involved 3 UCs describing a medical procedure at a Brazilian research institute. The aim was to describe how to apply a saline solution to skin areas with lesions. The quantity of solution and the adequate procedure to inject it on a patient depend on the location, appearance, and size of the lesions. The specification describes how to deal with each possible case. After the translation from the UCs to GT, 32 rules were created. In this case, the  evolution consisted in considering two types of lesions (small and large), removing different types of needles, and treating anesthetic buttons in an unified way. This evolution led to the merging of UCs (many rules became isomorphisms and were deleted, 2 UCs became equivalent -- i.e., they had repeated rules).

The bank case study contained 5 UCs describing how a customer accesses his account in a given bank through an ATM and how a bank employee can certify that a deposit is correct and, thus, can be completed. The shortest UC contained 7 steps, considering the main scenario and extensions, whereas the longest contained a total of 17 steps. The formalization resulted in a total of 48 rules. From this original GT, we applied evolution 1, described in the running example in Section~\ref{sec:hogt}, with an additional evolution rule to store the type of login performed. Based on this evolution, a non-predicted second evolution was necessary. The problem occurred because the login by card and password required a card to be inserted in the ATM, which must be dispensed when the costumer logs out. However, when the login was done by fingerprint, there was no card to be dispensed. Hence, the first evolution allowed a costumer to login by fingerprint but required a card to be dispensed at logout. The new second-order rule modified rules including a marker in situations where the login was by fingerprint. This rule induced inter-level conflicts, since the original rules were applicable to graphs without the marker, unlike the modified rules. This ensured the card must be dispensed at logout only if it was inserted at login, thus allowing the correct logout operation for each login type. 

The third case study considered the specification of an e-commerce application\footnote{Available at \url{http://www.utdallas.edu/~chung/RE/Presentations07S/Team\_3/UseCaseDocument.doc.} Last accessed on July 2, 2019.}. For this case study, we have formalized 5 UCs, considering the UC that corresponds to the login operation and, therefore, is required by all other UCs, and 4 other UCs related to basic operations of the system, such as browsing/searching the catalogue of items and dealing with account information. Each UC contained between 6 and 20 steps. The proposed evolution was to preserve the data of a user account when it is cancelled, rather than simply deleting the account. Hence, when cancelling an account, this account would just be marked as inactive and could be reactivated later on by a system manager. This forced other three necessary modifications: (i) all rules mentioning a user have to consider whether this user has an active account; (ii) when a new user account is created, it must be initialized as deactivated until the user confirms its activation; and (iii) all operations in the system must be restricted only to users with active accounts.  

\subsection{Results} \label{sub:results}

Table \ref{tab:comparison} shows the quantitative results of our case studies. It presents, for each case study: the number of UCs of each system; the total number of first-order rules present in the original GT; the  number of evolution rules created based on the proposed evolution;  number of inter-level conflicts (i.e., when a second-order rule creates a conflict with a first-order rule); the  number of rules affected by the evolution; the  number of rules deleted from the original GT because they became either obsolete (i.e., inapplicable) or useless (i.e., without any practical effect) after an evolution; and the  number of UCs modified as a consequence of the evolution.

\begin{table}[h]
  \centering
  \caption{Summary}

  \begin{tabular}{|l|c|c|c|}
  \hline
  \textbf{Case} & \textbf{Medical} & \textbf{Bank} & \textbf{Marvel} \tabularnewline \hline
  \# of UCs & 3 & 5 & 5\tabularnewline \hline 
  \# of 1st-order rules & 32 & 48 & 32 \tabularnewline \hline
  \# of 2nd-order rules & 8 & 2 & 4 \tabularnewline \hline
  \# of inter-level conflicts & 3 & 4 & 2 \tabularnewline \hline
  \# of affected rules & 11 & 7 & 5\tabularnewline \hline
  \# of deleted rules & 7 & 0 & 0 \tabularnewline \hline
  \# of affected UCs & 3 & 3 & 3 \tabularnewline \hline
  \end{tabular}

  \label{tab:comparison}
\end{table}

The number of inter-level conflicts indicates the points where special care has to be taken and decisions concerning the effects of an evolution have to be made. The sum of affected and deleted rules indicates the amount of work that would be necessary to apply the corresponding evolution. This means, for example, that applying the evolution of the Medical case study would require dealing with a total of 18 rules. However, we were able to automatically apply these changes by specifying 8 evolution rules. Considering that this specific case study contained 32 first-order rules, applying the evolution manually by identifying and modifying each rule could be difficult and error-prone. Instead, by specifying evolution rules, we can automatically detect all step rules that have to be changed, as well as which changes should be applied. 


We only considered 3 case studies with a few UCs but, even with this reduced number of cases, it was already possible to see the benefits our approach brings in contrast with modifications carried out by hand. In particular, the experiments showed how an evolution may trigger other changes, and keeping track of all these side-effects could be a hard task. Therefore, the experiments demonstrate our approach is useful, scalable, and applicable in practice.

Despite the need of some expertise to formally define the evolution, we noticed that often evolution rules are not complex to specify. Moreover, considering the real gain in having automatic analyses of the impact of system changes, it seems worth the effort. The interpretation of the analysis results also require some knowledge about conflicts and dependencies, but these concepts can be easily associated to similar ideas involving UCs. It is worth to mention that our work provides support for the analysis of the impacts of an intended evolution, but the user has the final decision on implementing it or not. As the user can analyse how the evolution affects the system, they can decide whether the impact is desirable and acceptable or the effects of the proposed evolution are more a problem than a solution.

\section{Related Work} \label{sec:relatedwork}

CPA has already been used to support evolution in scenarios not connected with UCs, such as described in \cite{DeLeenheer2008} and in \cite{Mens2007}. However, we consider that UC specification is important as a means to communicate with stakeholders. Although we specify evolution in the GT language, we propose the corresponding changes to be applied to the UCs so as to keep them up-to-date as a documentation of the software for developers and stakeholders.

Considering support for UC evolution, Rui and Butler \cite{Rui03,Rui07} discussed a metamodel for UCs along with a set of UC refactoring categories based on the metamodel. Refactorings correspond to changes in the syntax or structure of the UCs, such as changing or deleting a UC entity. Although the mechanics of some refactoring operations were discussed, the authors did not evaluate the effect of a UC refactoring in other UCs and offer no support for semantic changes. In \cite{Mantz2012}, an approach for evolution of metamodels described by graphs was presented. They consider how to change the metamodel and guarantee that models constructed from this metamodel are kept consistent. However, neither of the two aforementioned approaches considers a combination of UC evolution and GT to represent and analyze evolution. Although evolution rules work as a sort of metamodel, they are used to describe changes in behavior, rather than only describe restrictions on the connections of elements at the structural level.

In \cite{HajriGoknilBriandEtAl2018}, the authors present a tool-supported approach for analysing change impact regarding evolving configurations of product line use case models, which applies an idea similar to ours of impact analysis. They support decision-making about inclusion/exclusion of variant UCs, allowing the analysis of different configurations (i.e., sets and orders of UCs) for a product-line software and the effect of these changes in the specific products derived from this more general UC. A particular version of UC description is used to restrict rules and keywords constraining the use of natural language, thus enabling automatic changes in the product-specific (PS) software once product-line (PL) configuration changes have been approved. In this case, the PL UC plays a similar role as a second-order rules in our work and, due to the restrictions of the language used in the UCs, automated updates are possible. However, they focus on high-level decisions, changing configurations, whereas our approach deals with a fine-grain specification, concentrating on the specific behaviour described by each UC. Moreover, like other strategies used to automate the process of dealing with UCs, they restrict the use of natural language in the UC description. We believe this limits the expression of the requirements in terms of the end-user language and may preclude the expression of unforeseen scenarios. For this reason, we would like to keep the UC description in natural language and in the well-known format, but take advantage of the formal model. Furthermore, we also define an analysis of the impact of evolution on UCs even before their actual application. Therefore, we not only support the evolution of an existing system originated from a set of UCs, but also help the decision-making process related to whether implementing or not a certain evolution.

As far as we know, there is no previous work proposing the use of second-order graph grammars to describe evolution. We advocate that GG is an intuitive way of describing system behaviour and its mapping from UCs creates the possibility of formal analysis on natural-language specification and their modification.

\section{Conclusions} \label{sec:conclusions}
We described an approach to support the definition and analysis of impact of use case (UC) evolution, extending previous work on the formalization and analysis of UCs using graph transformation (GT). 
In this work, we employed a high-order graph transformation framework as basis for the definitions and analyses. This framework was extended to allow higher-order transformation of rules with NACs, since NACs are required when modelling conditional steps in use cases.
Analysis of the impacts of evolution is based on the conflicts between evolution rules and step rules. We detailed the conflicts that may occur and implemented this verification technique in the Verigraph tool. We also provided guidelines to interpret the results of this automatic verification, such that the developer can have a hint on possible sources of evolution errors and on how to correct them (Table \ref{fig:OIEvolution}).

We illustrated the impact of evolution within a single UC as a running example, but our approach may be used to support evolution of systems consisting of several UCs, where the analysis of the impacts of changes is an error-prone and time-consuming task. Results on experiments with larger systems were reported in Sect.~\ref{sec:experiments}. We noticed that even small changes usually affect more than one UC in a system, not always in obvious ways. Therefore, techniques to detect the impacts of an evolution in a system that can be automated, like the one proposed in this paper, are highly desirable.
Although we used the formal framework to provide a foundation for UC evolution, it is noteworthy that the formal model is generic and could be used as a semantic model for evolution of other artifacts used in the software development process (one would just have to define a translation from this artifact to GT).


As future work, we plan to analyse the impact of evolution rules on the CPA graph of a system, adding new hints on how behavior will be affected by evolution; and to work on test case generation and on improving the automation of  evolution by adding detection of isomorphic and duplicate rules.


\section*{References}
\bibliographystyle{elsarticle-num}
\bibliography{sbmf2019}

\end{document}